\documentclass{aa}
\usepackage{natbib}
\usepackage[varg]{txfonts}
\usepackage{graphicx}

\usepackage{booktabs}
\usepackage{multirow}
\usepackage{tabularx}
\usepackage{lscape}

\newcolumntype{L}[1]{>{\raggedright\arraybackslash}p{#1}}
\newcolumntype{C}[1]{>{\centering\arraybackslash}p{#1}}
\newcolumntype{R}[1]{>{\raggedleft\arraybackslash}p{#1}}

\usepackage{color} 

\def\Msun{\ifmmode{\mathrm M_\odot}\else{M$_\odot$}\fi}

\def\referee#1{{#1}}

\usepackage{tikz}
\newcommand*\circled[1]{\tikz[baseline=(char.base)]{
            \node[shape=circle,draw,inner sep=2pt] (char) {#1};}}
            
\begin{document}

\title{Dense gas is not enough: environmental variations in the star formation efficiency of dense molecular gas at 100\,pc scales in M51}


\author{M.~Querejeta\inst{1,2}, E.~Schinnerer\inst{3}, A.~Schruba\inst{4}, E.~Murphy\inst{5,6}, S.~Meidt\inst{3,7}, A.~Usero\inst{2}, A.~K.~Leroy\inst{8}, J.~Pety\inst{9,10}, F.~Bigiel\inst{11}, M.~Chevance\inst{12}, C.~M.~Faesi\inst{3}, M.~Gallagher\inst{8}, S.~Garc\'{i}a-Burillo\inst{2}, S.~C.~O.~Glover\inst{13}, A.~P.~S.~Hygate\inst{3,12}, M.~J.~Jim\'{e}nez-Donaire\inst{14}, J.~M.~D.~Kruijssen\inst{12,3}, E.~Momjian\inst{15}, E.~Rosolowsky\inst{16}, and D.~Utomo\inst{8}
}

\institute{European Southern Observatory, Karl-Schwarzschild-Stra{\ss}e 2, D-85748 Garching, Germany,  \email{mquereje@eso.org}
\and Observatorio Astron{\'o}mico Nacional (IGN), C/ Alfonso XII 3, E-28014 Madrid, Spain
\and  Max-Planck-Institut f\"{u}r Astronomie, K\"{o}nigstuhl 17, D-69117 Heidelberg, Germany
\and Max-Planck-Institut f\"{u}r extraterrestrische Physik, Giessenbachstra{\ss}e 1, D-85748 Garching, Germany
\ 
\and National Radio Astronomy Observatory, 520 Edgemont Road, Charlottesville, VA 22903, USA
\and Infrared Processing and Analysis Center, California Institute of Technology, MC 220-6, Pasadena, CA 91125, USA
\and Sterrenkundig Observatorium, Universiteit Gent, Krijgslaan 281 S9, B-9000 Gent, Belgium
\and Department of Astronomy, The Ohio State University, 140 West 18th Ave, Columbus, OH 43210, USA
\and IRAM, 300 rue de la Piscine, F-38406 Saint Martin d’H\`eres, France
\and Sorbonne Universit\'e, Observatoire de Paris, Universit\'e PSL, \'Ecole normale sup\'erieure, CNRS, LERMA, F-75005, Paris, France
\and Argelander-Institut für Astronomie, Universität Bonn, Auf dem Hügel 71, D-53121 Bonn, Germany
\and Astronomisches Rechen-Institut, Zentrum f\"{u}r Astronomie der Universit\"{a}t Heidelberg, M\"{o}nchhofstra{\ss}e 12-14, D-69120 Heidelberg, Germany
\and Instit\"ut  f\"{u}r Theoretische Astrophysik, Zentrum f\"{u}r Astronomie der Universit\"{a}t Heidelberg, Albert-Ueberle-Strasse 2, D-69120 Heidelberg, Germany
\and Harvard-Smithsonian Center for Astrophysics, 60 Garden St, Cambridge, MA 02138, USA
\and National Radio Astronomy Observatory, 1003 Lopezville Road, Socorro, NM 87801, USA 
\and Department of Physics, University of Alberta, Edmonton, AB T6G 2E1, Canada
}

\date{Received 18 December 2018 / Accepted 26 February 2019}

\abstract {It remains unclear what sets the efficiency with which molecular gas transforms into stars. Here we present a new VLA map of the spiral galaxy M51 in 33\,GHz radio continuum, an extinction-free tracer of star formation, at $3''$ scales (${\sim} 100$\,pc). We combined this map with interferometric PdBI/NOEMA observations of \mbox{CO(1-0)} and \mbox{HCN(1-0)} at matched resolution for three regions in M51 (central molecular ring, northern and southern spiral arm segments).
While our measurements roughly fall on the well-known correlation between total infrared and HCN luminosity, bridging the gap between Galactic and extragalactic observations, we find systematic offsets from that relation for different dynamical environments probed in M51; for example, the southern arm segment is more quiescent due to low star formation efficiency (SFE) of the dense gas, despite its high dense gas fraction.
\referee{Combining our results with measurements from the literature at 100\,pc scales,} we find that the SFE of the dense gas and the dense gas fraction anti-correlate and correlate, respectively, with the local stellar mass surface density. \referee{This is} consistent with previous kpc-scale studies.
In addition, we find a significant anti-correlation between the SFE and velocity dispersion of the dense gas.
Finally, we confirm that a correlation also holds between star formation rate surface density and the dense gas fraction, but it is not stronger than the correlation with dense gas surface density.
Our results are hard to reconcile with models relying on a universal gas density threshold for star formation and suggest that turbulence and galactic dynamics play a major role in setting how efficiently dense gas converts into stars.}


\titlerunning{Star formation efficiency in M51}
\authorrunning{M.~Querejeta et al.}

\maketitle 
\section{Introduction}
\label{Sec:introduction}

Studies within the Milky Way and in external galaxies show that there is a strong correlation between the presence of molecular gas and star formation \citep[e.g.][]{2008AJ....136.2846B,2011AJ....142...37S,2013AJ....146...19L}, and this correlation seems to become more linear for observations tracing dense regions of molecular clouds, e.g.\ in HCN emission \citep{2004ApJS..152...63G,2004ApJ...606..271G,2005ApJ...635L.173W, 2010ApJ...724..687L, 2012A&A...539A...8G, 2015ApJ...805...31L}. However, the causal link between the two observables is not clear; in fact, models in which stars form with a constant depletion time when gas exceeds a certain density threshold have been questioned by recent observational studies \citep[][]{2013MNRAS.429..987L,2014MNRAS.440.3370K,2015AJ....150..115U,2016ApJ...822L..26B,2018ApJ...858...90G}. Along the same lines, observations of giant molecular clouds (GMCs) in external galaxies have highlighted the importance of dynamical environment in preventing molecular gas from collapsing, limiting its ability to form stars \citep{2013ApJ...779...45M,2017ApJ...839..133P,schruba19}.
In M51, the local dynamical state of the gas, assessed by its virial parameter, also appears to correlate with the ability of gas to form stars, with more strongly self-gravitating gas forming stars at a higher normalised rate \citep{2017ApJ...846...71L}.

So far, most studies comparing dense gas and star formation in external galaxies have focussed on integrated or relatively low-resolution (${\gtrsim} 1.5$~kpc) measurements  \citep[e.g.][]{2004ApJS..152...63G,2012A&A...539A...8G,2015AJ....150..115U,2016ApJ...822L..26B}. High-resolution maps of dense gas are still rare, although there are some exceptions \citep[e.g.][]{2015ApJ...813..118M,2015ApJ...815..103B,2017ApJ...836..101C,2018ApJ...862..120K,2018MNRAS.475.5550V}. 
A second issue affecting these studies is the difficulty to trace star formation {\em locally} in a way that is robust to extinction by dust. In this paper, we address both issues, presenting new high-resolution (${\sim} 100$~pc) maps of the dense gas tracer HCN and recent star formation traced by 33\,GHz radio continuum in M51. Combined with the PAWS CO survey \citep{2013ApJ...779...42S}, these new datasets allow us to make one of the first highly resolved comparisons between bulk molecular gas, dense gas, and recent star formation in several regions of a nearby galaxy.

Star formation rate (SFR) estimates based on broad-band continuum emission either assume that all light comes from a young stellar population or require uncertain corrections for infrared `cirrus' emission due to heating by old stars \citep[e.g.][]{2011ApJ...735...63L,2012MNRAS.426..892G,2012AJ....144....3L}. Moreover, given that stellar populations can emit low-level UV light for a long time, tracers like the total infrared (TIR) and UV are sensitive to the assumed star formation history, with degeneracies between the age and mass of stars formed. Tracers based on ionising photons such as recombination lines or free-free emission avoid this ambiguity, capturing light from clearly young populations \citep{2012ARA&A..50..531K}. They do so at the expense of being sensitive only to the upper end of the mass function, with potential concerns for IMF variations and stochasticity in the sampling; this applies both to SFRs estimated from 33\,GHz radio continuum and to tracers such as H$\alpha$.

A main obstacle to using recombination lines, such as H$\alpha$, to trace the SFR is that they are often significantly affected by dust extinction \citep[e.g.][]{2007ApJ...666..870C}. The radio free-free emission avoids this concern, and in this sense represents a powerful alternative to challenging observations of near-IR recombination lines \citep{2005ApJ...633..871C,2007ApJ...671..333K}.
Over ${\sim} 30-150$\,GHz, free-free emission should dominate radio continuum emission from a star-forming galaxy \citep{1992ARA&A..30..575C}. Since free-free emission emerges from the random, close encounters between charged particles (typically, an ion and a free electron), it is excellent in tracing star-forming \ion{H}{II} regions, with its flux being directly proportional to the production of ionising photons from newborn stars.
For a fully sampled IMF, in turn, the ionising photon production rate is proportional to the rate of recent star formation \citep[e.g.][]{1998ARA&A..36..189K,2012ARA&A..50..531K}. Because emission at these frequencies is almost totally unaffected by dust, the free-free continuum has been proposed as an optimal SFR tracer \citep[e.g.][]{1967ApJ...147..471M,1983ApJ...268L..79T,1988A&A...190...41K,2011ApJ...737...67M,2012ApJ...761...97M,2014ApJ...780...19R}.

So far, most exploration of the free-free continuum in normal star-forming galaxies has focussed on pointed observations of individual regions \citep[e.g.][]{2011ApJ...737...67M,2012ApJ...761...97M,2014ApJ...780...19R} or observations at lower frequencies, where separating the synchrotron and free-free emission represents a major concern \citep{1997A&A...322...19N}. With the upgrade to the continuum sensitivity of the Karl G. Jansky Very Large Array (VLA), it is now possible to map large parts of a galaxy at good resolution and sensitivity. In this paper, we present the first wide-area, highly resolved ($3'' \sim 100$~pc) map of 33~GHz emission across the nearby spiral galaxy M51. Pairing this map with new NOEMA observations of the high critical density rotational transition \mbox{HCN(1-0)} and the PAWS \mbox{CO(1-0)} survey, we can compare dense gas, total molecular gas, and extinction-free estimates of recent star formation across the inner star-forming part of M51.

In Sect.~\ref{Sec:data} we describe the observations and how we perform aperture photometry.
Sect.~\ref{Sec:SF-gas} presents the main results of the paper, starting from the relation between TIR luminosity and HCN emission (Sect.~\ref{Sec:HCN-TIR}), highlighting some spatial offsets between tracers (Sect.~\ref{Sec:offsets}), quantifying variations of star formation efficiency among the regions that we target (Sect.~\ref{Sec:SFE}), and testing the predictability of SFR surface density based on the dense gas fraction (Sect.~\ref{Sec:Viaene}). We summarise the limitations and caveats inherent to the present study in Sect.\,\ref{Sec:caveats}. We discuss the implications of our results in Sect.~\ref{Sec:discussion}, and we close with a short summary of the main conclusions in Sect.~\ref{Sec:concl}.

Throughout the paper, we assume a distance of $D=8.58$\,Mpc to M51 \citep[which has a statistical uncertainty of only 0.1\,Mpc;][]{2016ApJ...826...21M}; this implies that $3''=125$\,pc. For simplicity, we will refer to our results as ``100\,pc scale'' measurements, as opposed to the ${\sim}$kpc-scale measurements previously available from single dish observations. We assume an inclination of $i=22^\circ$ for the disc of M51, which is well constrained from kinematics by PAWS \citep{2014ApJ...784....4C}.

\section{Observations and data reduction} 
\label{Sec:data}

\subsection{33\,GHz continuum from the VLA} 
\label{Sec:33GHz}

\begin{figure*}[t]
\begin{center}
\includegraphics[trim=10 180 70 150, clip,width=0.95\textwidth]{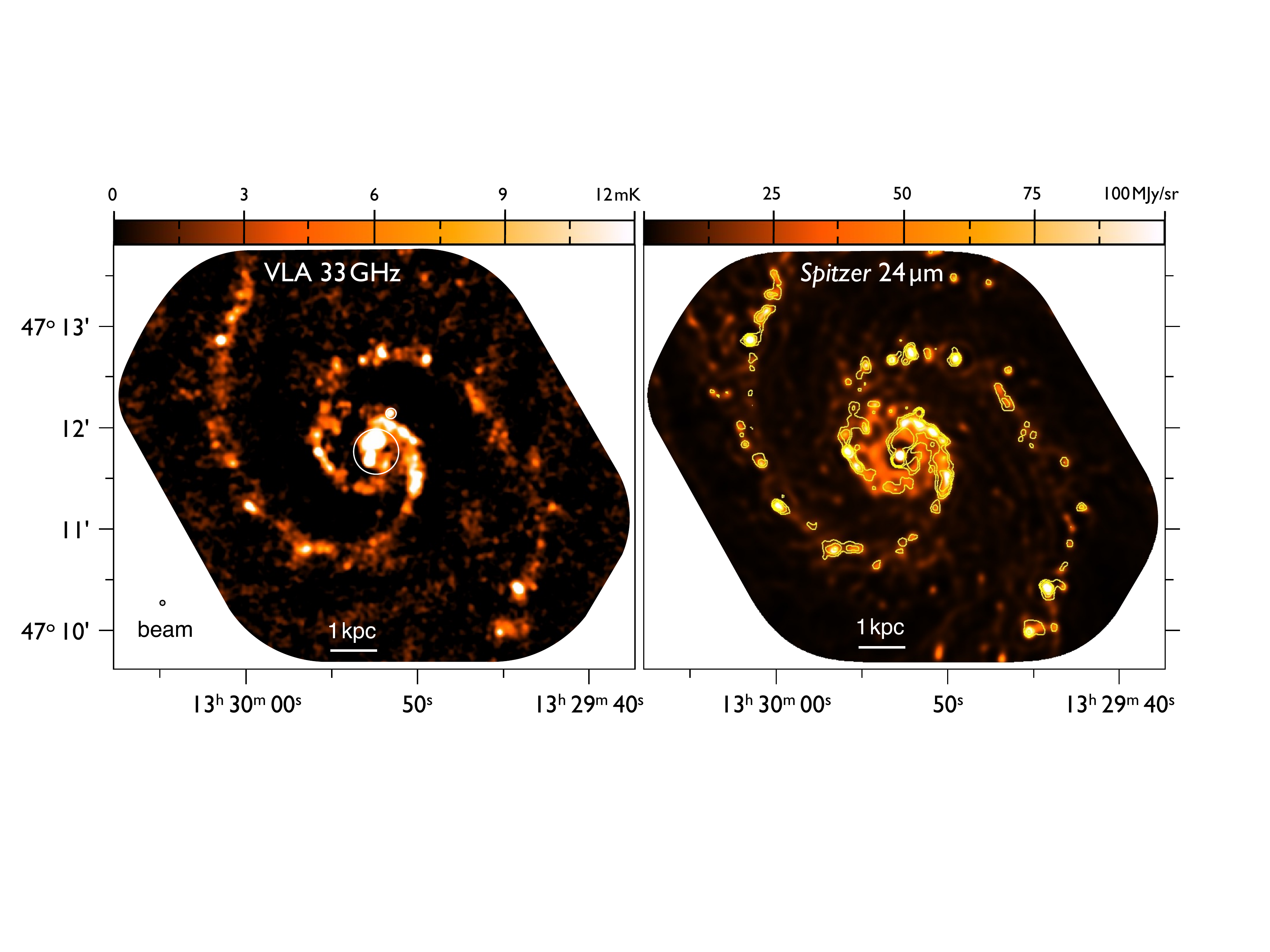}
\end{center}
\caption{\textit{Left panel:} 33\,GHz continuum map of M51 from the VLA at $3''$ (125\,pc) resolution. The white circles indicate the area affected by the AGN, which we exclude from the analysis. 
\textit{Right panel:} \textit{Spitzer}/MIPS 24\,$\mu$m image from \citet{2011AJ....141...41D}, who deconvolved the PSF, resulting in a map of comparable resolution. The overplotted yellow contours are from the 33\,GHz map (levels [3,6,12]\,mK). The intensity scales of both maps correspond to similar SFR values. The field of view is $300'' \times 250''$ (${\sim} 11 \times 9$\,kpc), with the horizontal axis corresponding to right ascension (east to the left) and the vertical axis to declination (north up), relative to J2000.
}
\label{fig:VLA-24um}
\end{figure*}

We observed M51 with the Karl G. Jansky Very Large Array (VLA) using the Ka-band ($26.5{-}40$\,GHz) receiver. We used the 3-bit samplers that delivered four baseband pairs of 2 GHz each, in both right- and left-hand circular polarisation; they were centred at 30, 32, 34, and 36\,GHz, with a total simultaneous bandwidth of 8\,GHz. These observations were carried out between August 2014 and January 2015 for a total of 50\,h including overheads (project VLA/14A-171). Out of these, 44\,h were observed in C-configuration ($b_{\rm max} = 3.4$\,km, corresponding to an angular scale of $0.6''$ in the Ka-band), while 6\,h were added in the most compact D-configuration ($b_{\rm max} = 1$\,km, angular scale of $2''$). The galaxy was covered using a hexagonally-packed Nyquist-sampled mosaic of 20 pointings. Given the VLA configurations employed, our map is sensitive to spatial scales up to $44'' {\sim} 1.6$\,kpc. We discuss the potential of resolving out emission due to the missing short spacings and some other caveats in Sect.~\ref{Sec:usefultracer}.

We reduced the VLA observations with the  Common  Astronomy  Software Applications \citep[CASA;][]{2007ASPC..376..127M}. Specifically, we closely followed the standard VLA pipeline, inspecting results for each dataset carefully after initial calibration, editing problematic data, and running a new instance of the pipeline with those problematic data flagged. Then, we concatenated all calibrated datasets into a single measurement set, and imaged it using the Briggs weighting scheme within the task \texttt{tclean} in CASA (in multi-frequency synthesis mode, Briggs robustness parameter 0.4). While imaging with multi-scale cleaning, we assumed a flat spectrum over the full frequency range $29{-}37$\,GHz, and we truncated the map where the primary beam response falls below ${<} 0.4$ of the peak (i.e.\ avoiding the edges).
We deconvolved the map down to a threshold of $12\,\mu$Jy/beam. In order to increase the signal-to-noise ratio of the resulting image and to match the spatial resolution to the HCN datasets, our final 33\,GHz map was imaged with a taper that results in a round beam of $3'' \times 3''$ (while the original resolution in natural weighting was $0.82'' \times 0.75''$; we will use this version in Sect.\,\ref{Sec:offsets}). We estimate the uncertainty on our aperture measurements by empirically measuring the rms noise on emission-free areas of the 33\,GHz image, and adding in quadrature the standard VLA flux calibration uncertainty \citep[${\sim}$3\% for Ka-band;][]{2013ApJS..204...19P}. The rms noise of our final map before primary-beam correction is ${\sim}6\,\mu$Jy/beam (relative to the $3'' \times 3''$ beam and before adding the calibration uncertainty in quadrature). The conversion factor from Jy/beam to K is 124.7.

Figure~\ref{fig:VLA-24um} shows our map of 33\,GHz radio continuum; for illustrative purposes, we also show the \textit{Spitzer} 24\,$\mu$m image from \citet{2011AJ....141...41D}. We choose intensity scales such that the brightest regions in the 33\,GHz and 24\,$\mu$m maps translate into similar SFR values (following \citealt{2011ApJ...737...67M} and \citealt{2007ApJ...666..870C}, respectively).
We note that the 24\,$\mu$m image has comparable resolution to our 33\,GHz map ($3'' \approx 100$\,pc) as the result of a deconvolution of the \textit{Spitzer} MIPS point spread function (PSF) presented in \citet{2011AJ....141...41D}. There is good visual agreement with 24\,$\mu$m, but not a perfect one-to-one match, as we discuss in Sect.~\ref{Sec:usefultracer} and Appendix~\ref{sec:SFRcomparison}.

\subsection{HCN emission from NOEMA} 
\label{Sec:densedata}

We targeted the \mbox{HCN(1-0)} transition at $\nu_\mathrm{rest} = 88.6$\,GHz in three different areas of M51 using the IRAM PdBI/NOEMA interferometer. We corrected for missing short spacings by combining our data with observations from the IRAM 30\,m telescope for the same frequencies \citep{2016ApJ...822L..26B}. Our final maps have a spatial resolution of ${\sim} 3''$, with channels of 2.07\,MHz ($\approx$7\,km/s).
Fig.~\ref{fig:regions} shows the areas that we have mapped. We will refer to the three distinct areas as ``ring'' (centred at the nucleus, RA=13:29:52.708, Dec=+47:11:42.81), ``north'' (centred on the spurs of the northern arm, RA=13:29:50.824, Dec=+47:12:38.83), and ``south'' (south-western arm segment, RA=13:29:51.537, Dec=+47:11:01.48). All coordinates in this paper are equatorial relative to J2000.0. The primary beam of NOEMA at the HCN(1–0) frequency is $56.9''$ (FWHM), and Fig.~\ref{fig:regions} shows the position of our fields of view (for reference, the large circles indicate a truncation radius of $R=35''$, beyond which we do not consider any apertures).

For the nuclear pointing, the PdBI/NOEMA observations were carried out between November 2011 and March 2016, covering all four configurations of the interferometer (A, B, C, and D). The southern pointing was observed between July 2014 and November 2015, in configurations C and D.  The northern pointing is actually a 2-pointing mosaic, observed between February and April 2015 in B configuration. 
For most of the observations we used the source MWC349 as our flux calibrator, and we expect an uncertainty of ${\sim} 5$\% for the absolute flux calibration in these cases (M.~Krips, private communication). We expect slightly higher uncertainties for other flux calibrators (e.g.\ LkHa101), up to ${\sim} 10$\%, which we had to rely on for a few days when MWC349 was not available.  Thus, to construct uncertainty maps we assume a conservative flux calibration uncertainty of ${\sim} 10$\%, which we add in quadrature to the rms noise measured on line-free channels.
    
We employed the wide-band correlator (WideX) and the narrow-band correlator simultaneously. This allowed us to map \mbox{HCN(1-0)}, \mbox{HNC(1-0)}, and \mbox{HCO$^+$(1-0)} at the same time, but in this paper we focus only on HCN.
We calibrated and mapped our observations with GILDAS \citep{2005sf2a.conf..721P}. The minimum separation between the PdBI/NOEMA antennas is ${\sim}24$\,m, which means that, for the frequency of \mbox{HCN(1-0)}, angular scales larger than ${\sim} 35''$ will be filtered out by the interferometer even in the most compact configurations. To overcome this problem, we corrected for missing short spacings with the task \texttt{uv\_short} within GILDAS, using the single-dish observations from the EMPIRE survey \citep{2016ApJ...822L..26B}. These data were obtained with the 3\,mm-band EMIR receiver on the IRAM 30\,m telescope under typical summer conditions in July-August 2012, and calibrated with CLASS. The native spatial resolution is ${\sim} 28''$ (in practice, the data were convolved to ${\sim} 30''$ resolution), and the spectral axis was regridded from a native resolution of 195\,kHz to 7\,km\,s$^{-1}$ channels. We refer the reader to \citet{2016ApJ...822L..26B} for additional details.

We used the Hogbom algorithm for cleaning, initially with natural weighting. The nuclear pointing results in a higher spatial resolution (a synthesised beam of $2.7'' \times 1.9''$) than the other two areas (${\sim} 3''$ resolution). This is why, in a second step, we imaged all three cubes with a taper (robust weighting) to produce maps of matched spatial resolution with a circular synthesised beam of ${\rm FWHM} = 3.0''$.  
We also performed an alternative reduction of the nuclear pointing at higher resolution, which we use in Sect.\,\ref{Sec:offsets}. We imaged the nuclear HCN pointing using robust weighting (robust parameter\,$=10.0$ in GILDAS), which results in a synthesised beam of $1.58'' \times 1.18''$.
We have applied a primary beam correction to our maps (primary beam of FWHM=$56.9''$), truncating at $R=35''$ (where the sensitivity drops below 35\% of the peak for a single pointing; these are the coloured circles shown in Fig.~\ref{fig:regions}). Since the HCN linewidths that we measure are typically larger than $\mathrm{FWHM} \gtrsim 20$\,km/s, we ignore any hyperfine-structure details of the HCN line. 
The two strongest hyperfine components of HCN are separated by only ${\sim} 5$\,km/s, and the maximum separation between (weaker) hyperfine components is ${\sim} 12$\,km/s; we have estimated that hyperfine splitting can affect our linewidth measurements at most at the ${\lesssim}10$\% level.
The average 1$\sigma$ noise level in the central $20''$ (diameter) of each pointing
is 22\,mK for the northern pointing and 12\,mK for the southern pointing (over 7\,km\,s$^{-1}$ channels and measured on the primary beam-corrected cubes). For the ring area, we have an average 1$\sigma$ noise of 12\,mK in an annulus extending between $R=15''$ and $R=25''$ (we exclude a central circle from the analysis to avoid contamination from the AGN; see Sect.~\ref{Sec:SFRs}).

We calculate the total intensity of CO and HCN emission as a function of position by integrating over the cube excluding line-free channels. Specifically, we integrate in the range (367, 451)\,km/s for the northern pointing, (346, 591)\,km/s for the central ring, and (437, 598)\,km/s for the southern pointing (all quoted velocities are relative to the local standard of rest); the systemic velocity of the galaxy is $v_{\rm sys} \approx 472$\,km/s \citep{2014ApJ...784....4C}. 
This is the traditional zeroth-order moment map without applying any thresholds (shown in Fig.~\ref{fig:regions}).  
We estimate the uncertainty of moment zero maps as $\sigma_\mathrm{line} = \mathrm{rms}_\mathrm{channel} \times \delta V \times \sqrt{N_\mathrm{window}}$, where $\mathrm{rms}_\mathrm{channel}$ is the pixel-by-pixel rms noise computed from line-free channels (in Kelvin), $\delta V$ is the channel width (in km\,s$^{-1}$), and $N_\mathrm{window}$ is the number of channels used in the integration window. To this uncertainty we add in quadrature the typical flux calibration uncertainty of $10\%$.

We measure the velocity dispersion in the emission lines, $\sigma$, following the ``effective width'' approach \citep{2001ApJ...551..852H,2016ApJ...831...16L,2017ApJ...846...71L,2018ApJ...860..172S}:

\begin{equation} \label{eq:linewidth}
\sigma = \frac{I_\mathrm{tot}}{T_\mathrm{peak} \sqrt{2 \pi}},
\end{equation} 

\noindent
where $I_\mathrm{tot}$ is the total integrated intensity (K\,km\,s$^{-1}$) and $T_\mathrm{peak}$ is the peak brightness temperature in the spectrum (K). This method is robust to the presence of noise and line wings, but it can potentially be biased, for example, by noise spikes. We consider pixels above $5 \times \mathrm{rms}$ to retain only the sight-lines with high significance.

To confirm that these velocity dispersion measurements are meaningful, we also constructed second moment maps using a variant of the window method \citep{1981AJ.....86.1791B}. This technique can capture more low-level emission than moment maps based on simple noise clipping (which removes the wings below the imposed threshold), and thus can provide more robust estimates of the velocity dispersion. We first identify pixels with significant detections using standard sigma-clipping (we consider pixels with signal above $5 \times \mathrm{rms}$, while the remaining pixels are blanked); then, for each significant pixel, we define a range of channels (the frequency ``window'') where emission will be considered to construct moment maps. This window is defined by iteratively expanding the range of channels, starting from the peak channel (the maximum flux of the spectrum), until the average continuum outside the window converges to the criterion from \citet{1981AJ.....86.1791B}.
We compute uncertainty maps for the velocity dispersion by formally propagating $\mathrm{rms}_\mathrm{channel}$ through the relevant equations. The smallest linewidths that we can reliably measure are $\sigma \gtrsim 10$\,km\,s$^{-1}$ (or ${\rm FWHM} \gtrsim 23$\,km\,s$^{-1}$), so we do not expect a significant bias due to the finite channel width (since our channels are 7\,km\,s$^{-1}$ wide; see \citealt{2016ApJ...831...16L}).

\subsection{CO molecular gas emission} 
\label{Sec:COdata}

We also used the \mbox{CO(1-0)} moment maps from PAWS \citep{2013ApJ...779...42S} tapered to $3''$ resolution as a tracer of bulk molecular gas. The PAWS maps were corrected for missing short-spacings with IRAM 30\,m single-dish data, as described in \citet{2013ApJ...779...43P}, where the interested reader can find additional information on calibration and imaging.

\subsection{Stellar mass surface density} 
\label{Sec:s4gdata}

We use a stellar mass map based on \textit{Spitzer} 3.6\,$\mu$m imaging, from the \textit{Spitzer} Survey of Stellar Structure in Galaxies \citep[S$^4$G;][]{2010PASP..122.1397S}. The 3.6\,$\mu$m image has been corrected for dust emission using a pipeline that applies Independent Component Analysis (ICA) to the adjacent \textit{Spitzer} 3.6\,$\mu$m and 4.5\,$\mu$m  bands \citep{2012ApJ...744...17M,2015ApJS..219....5Q}; the different spectral energy distribution of stars and dust for these wavelengths permits us to separate two components (stars and dust) on a pixel-by-pixel basis. 
In a second step, the stellar component isolated by ICA is transformed to stellar mass surface density with a constant mass-to-light ratio of 0.6\,$M_\odot/L_\odot$, which \citet{2014ApJ...788..144M} argue provides an accurate conversion between dust-free stellar emission and stellar mass at 3.6\,$\mu$m \citep[in agreement with][]{2014ApJ...797...55N,2015MNRAS.449.2853R}. The map that we use here includes some small improvements relative to the map of M51 publicly released\footnote{http://irsa.ipac.caltech.edu} in 2015, as it uses a more effective way of interpolating over regions strongly dominated by dust emission. The assumption of two components with ICA can result in unphysical local flux depressions in the stellar mass map \citep{2015ApJS..219....5Q,2016A&A...588A..33Q}; we identify areas with flux significantly below the azimuthal average at each radius, and fill them with the average flux from the immediate surroundings, imposing random scatter representative of photonic noise. In any case, these regions represent a small fraction of the total area of the galaxy. Our map shows good agreement with the stellar mass map for M51 obtained by \citet{2017ApJ...835...93M} using Bayesian marginalisation analysis.

\subsection{Multi-wavelength radio data} 
\label{Sec:multiradio}

To calculate thermal fractions at 33\,GHz radio continuum to estimate SFR in our apertures, we also use archival 8.4, 4.9, and 1.4\,GHz maps from the VLA (3.6, 6, and 20\,cm). They were presented in \citet{2011AJ....141...41D}, and have a resolution of $2.4''$, $2.0''$, and $1.4''$. They achieved an rms of 25, 16, and 11\,$\mu$Jy/beam, respectively.

\subsection{Spitzer 24\,$\mu$m image} 
\label{Sec:24um}

For comparison purposes, we use a \textit{Spitzer} 24\,$\mu$m map of M51 as an alternative SFR tracer. This map comes from \citet{2011AJ....141...41D}, and it has an enhanced resolution of $2.3'' \times 2.4''$
because the authors applied an algorithm to deconvolve the instrumental PSF \citep[HiRes;][]{2005ASPC..347...61B}. We note that no diffuse ``cirrus'' emission has been removed (i.e.\ a component associated with heating by older stars).

\subsection{Conversion to physical parameters} 
\label{Sec:SFRmap}

\subsubsection{Star formation rates} 
\label{Sec:SFRs}

Our new observations of the 33 GHz continuum emission from M51 with the VLA allow us to estimate SFRs at ${\sim} 100$\,pc scales. We first determine the fraction of free-free emission at 33\,GHz for each of the apertures (the ``thermal fraction'') and then convert the thermal fluxes to SFRs.

\paragraph{Determination of the free-free fraction:} 
\label{Sec:freefree}

A good approximation of the thermal fraction can be obtained measuring the spectral index between 33\,GHz and some neighbouring radio bands, assuming a fixed spectral slope for synchrotron emission. We have followed this approach, using Eq.~(11) from \citet{2012ApJ...761...97M}, which we reproduce next as Eq.~(\ref{eq:tfrac}); this estimation of the thermal fraction, $f_{\rm T}^{\nu_1}$,
relies on a thermal radio spectral index of 0.1 (optically thin free-free emission):

\begin{equation} \label{eq:tfrac}
f_{\rm T}^{\nu_1} = \frac  {  \left(\frac{\nu_2}{\nu_1}\right)^{-\alpha}   -  \left(\frac{\nu_2}{\nu_1}\right)^{-\alpha^\mathrm{NT}}  }      {  \left(\frac{\nu_2}{\nu_1}\right)^{-0.1}   -  \left(\frac{\nu_2}{\nu_1}\right)^{-\alpha^\mathrm{NT}}   },
\end{equation} 

\noindent
where, for our case, $\nu_1 = 33$\,GHz, $\nu_2 = 1.4$\,GHz, and the measured spectral index, $\alpha$, is obtained through a least-squares fit that simultaneously considers multiple radio bands,  8.4\,GHz, 4.9\,GHz, and 1.4\,GHz \citep[from][]{2011AJ....141...41D}. We set $\alpha^\mathrm{NT} = 0.85$, in agreement with the measurements of \citet{2011ApJ...737...67M} in NGC\,6946, and similar to the average value of $\alpha^\mathrm{NT} = 0.83$ from \citet{1997A&A...322...19N}. The spectral index $\alpha$ is calculated for each of our apertures after convolving the maps to the common $3''$ resolution, and taking the relevant photometric uncertainties into account. We also consider an uncertainty of $\Delta \alpha^\mathrm{NT} = 0.06$ for the non-thermal spectral index, which is the typical uncertainty in $\alpha^\mathrm{NT}$ found by \citet{2011ApJ...737...67M} among ten individual regions in NGC\,6946.
Applying Eq.~(\ref{eq:tfrac}), we obtain a thermal fraction for each aperture with an associated uncertainty (quoted in Table~\ref{table:results}). 

The median thermal fraction for significant detections on apertures centred on HCN peaks is $f_{\rm T}=64 \pm 14$\%. The thermal fraction becomes slightly higher for apertures placed on 33\,GHz peaks (median $f_{\rm T}=72 \pm 11$\%) and slightly lower for all significant detections on Nyquist-sampled apertures (median $f_{\rm T}=56 \pm 16$\%). These differences might be expected as free-free emission on 100\,pc scales contributes more towards bright star-forming regions (33\,GHz peaks). Our thermal fractions are slightly lower than the average ${\approx} 76$\% found by \citet{2012ApJ...761...97M}, with a dispersion of $24$\%, or the average values in the range 70${-}$85\% found by \citet{2015ApJ...813..118M}; they are also lower than the typical 80${-}$90\% thermal fractions found among 112 regions at ${\approx}30{-}300$\,pc resolution within the Star Formation in Radio Survey (Linden et al.\ in prep.).

The relatively low thermal fractions that we find in M51 might be associated with the fact that M51 is an ongoing merger \citep[lower thermal fractions at 33\,GHz have been measured in the centre of some local (U)LIRGs; see e.g.][]{2013ApJ...777...58M,2015ApJ...799...10B}. Moreover, given that here we only focus on regions within the inner ${\lesssim}3$\,kpc of the galaxy, the Seyfert 2 nucleus of M51 (which hosts a radio jet) might also be contributing diffuse synchrotron emission to our apertures. One would expect this effect to be more pronounced for the apertures that are closest to the nucleus, and this is indeed what we find: the thermal fraction is higher for the northern region that we target (at $R_\mathrm{gal}\sim 2.5$\,kpc), with a median $f_{\rm T}=90 \pm 5$\% for apertures on 33\,GHz peaks, as opposed to $f_{\rm T}=71 \pm 10$\% for analogous apertures in the ring ($R_\mathrm{gal}\sim 1$\,kpc).
In any case, we emphasise that for the calculation of SFRs we adopt the specific thermal fraction determined for each aperture.

\paragraph{Conversion of thermal fluxes to SFRs:} 
\label{Sec:fluxtoSFR}

In converting the thermal 33\,GHz fluxes into SFRs, we follow Eq.~(11) from \citet{2011ApJ...737...67M}:

\begin{equation} \label{eq:conversionSFR}
\left(\frac{\mathrm{SFR}_\nu^\mathrm{T}}{M_\odot \, \mathrm{yr^{-1}}}\right) = 4.6 \times 10^{-28} \left(\frac{T_{\rm e}}{10^4\,\mathrm{K}}\right)^{-0.45}  \left(\frac{\nu}{\mathrm{GHz}}\right)^{0.1} \left(\frac{L_\nu^\mathrm{T}}{\mathrm{erg\,s^{-1}\,Hz^{-1}}}\right),
\end{equation}

\noindent
where $T_{\rm e}$ is the electron temperature, $\nu$ is the frequency (in our case, 33\,GHz), and $L_\nu^\mathrm{T}$ is the luminosity of free-free emission at 33\,GHz (i.e.\ the \textit{thermal} contribution only, $L_\nu^\mathrm{T}=f_{\rm T}\,L_\nu$).
Fortunately, electron temperatures are observationally well constrained for M51, and we adopt $T_{\rm e} = (6300 \pm 500)$\,K from \citet{2015ApJ...808...42C}, which is in good agreement with \citet{2004ApJ...615..228B}. Thus, and given the mild dependence of SFR on electron temperature (through the ${-}0.45$ exponent), even an extreme variation of $\pm 900$\,K in $T_{\rm e}$ has an impact of at most 7\% on the SFR. However, there can be additional systematic uncertainties regarding the conversion of 33\,GHz to SFRs, and we discuss them in Sect.\,\ref{Sec:usefultracer}.

M51 has a well-studied AGN with a radio plasma jet \citep{1992AJ....103.1146C,2016A&A...593A.118Q}. 
This very central area is clearly dominated by synchrotron emission (as evidenced by the strong morphological similarity between 33\,GHz and the 20\,cm radio continuum maps), and we exclude it from our analysis.
 Therefore, in all maps for the ``ring'' area we exclude a circle of radius $R=13.4''$ (centred on RA=13:29:52.432, Dec=+47:11:44.85). As shown by \citet{2015MNRAS.452...32R}, the bright and compact radio feature $27.8''$ north of the nucleus is aligned with the nuclear radio jet and is in all likelihood associated with past nuclear activity; thus, we also exclude from our analysis this component as a circle of radius $R=3.2''$ centred on RA=13:29:51.585, Dec=+47:12:07.61. These areas are indicated as white circles on the left panel of Fig.\,\ref{fig:VLA-24um}.

\subsubsection{Molecular gas surface densities} 
\label{Sec:gassurfdens}

With our molecular emission line tracers, \mbox{HCN(1-0)} and \mbox{CO(1-0)}, we are sensitive to different phases of the molecular interstellar medium, with HCN tracing gas at higher densities (as we will further discuss in Sect.~\ref{Sec:HCNproblems}). 
To convert \mbox{HCN(1-0)} fluxes into (dense) molecular gas surface densities, we assume the standard conversion factor $\alpha_\mathrm{HCN}  = 10\,M_\odot$\,(K\,km\,s$^{-1}$\,pc$^2)^{-1}$, suggested by \citet{2004ApJS..152...63G} for gas above a density of $n \approx 6.0 \times 10^4$\,cm$^{-3}$, which includes a correction for helium. We note that \citet{2018MNRAS.479.1702O} recently derived a higher value using numerical simulations, $\alpha_\mathrm{HCN}  = 14\,M_\odot$\,(K\,km\,s$^{-1}$\,pc$^2)^{-1}$ (for gas above a density of $n \approx 1.0 \times 10^4$\,cm$^{-3}$); however, since we aim to compare to previous measurements and trends from the literature, we prefer to adopt $\alpha_\mathrm{HCN}  = 10\,M_\odot$\,(K\,km\,s$^{-1}$\,pc$^2)^{-1}$. Given the shallow metallicity gradient in M51 \citep{2004ApJ...615..228B,2010ApJS..190..233M,2015ApJ...808...42C}, and the fact that we avoid the nuclear area, where the conversion factor is expected to be lower \citep{2016A&A...593A.118Q}, the standard value is probably a reasonable assumption. Yet, we warn the reader that a number of effects can lead to variations in $\alpha_\mathrm{HCN}$ \citep{2017ApJ...835..217L}, yielding dense gas masses that are somewhat uncertain.

We use the standard (Galactic) conversion factor of $\alpha_\mathrm{CO} = 4.4\,M_\odot$\,(K\,km\,s$^{-1}$\,pc$^2)^{-1}$, which includes a correction for heavy elements, to transform the measured \mbox{CO(1-0)} intensities into molecular gas surface density \citep{2013ARA&A..51..207B}. Since metallicity gradients are shallow in M51, 
and we exclude the area around the nucleus, we do not expect large variations in $\alpha_\mathrm{CO}$ \citep{2010ApJ...719.1588S,2013ApJ...777....5S}. \citet{2017ApJ...835..217L} solved for $\alpha_\mathrm{CO}$ in M51 using dust emission (see their Appendix A2), which quantitatively motivates the choice of the standard conversion factor for the regions that we have targeted.

\subsection{Aperture photometry} 
\label{Sec:aperturephoto}

\begin{figure*}[t]
\begin{center}
\includegraphics[trim=80 20 170 0, clip,width=1.0\textwidth]{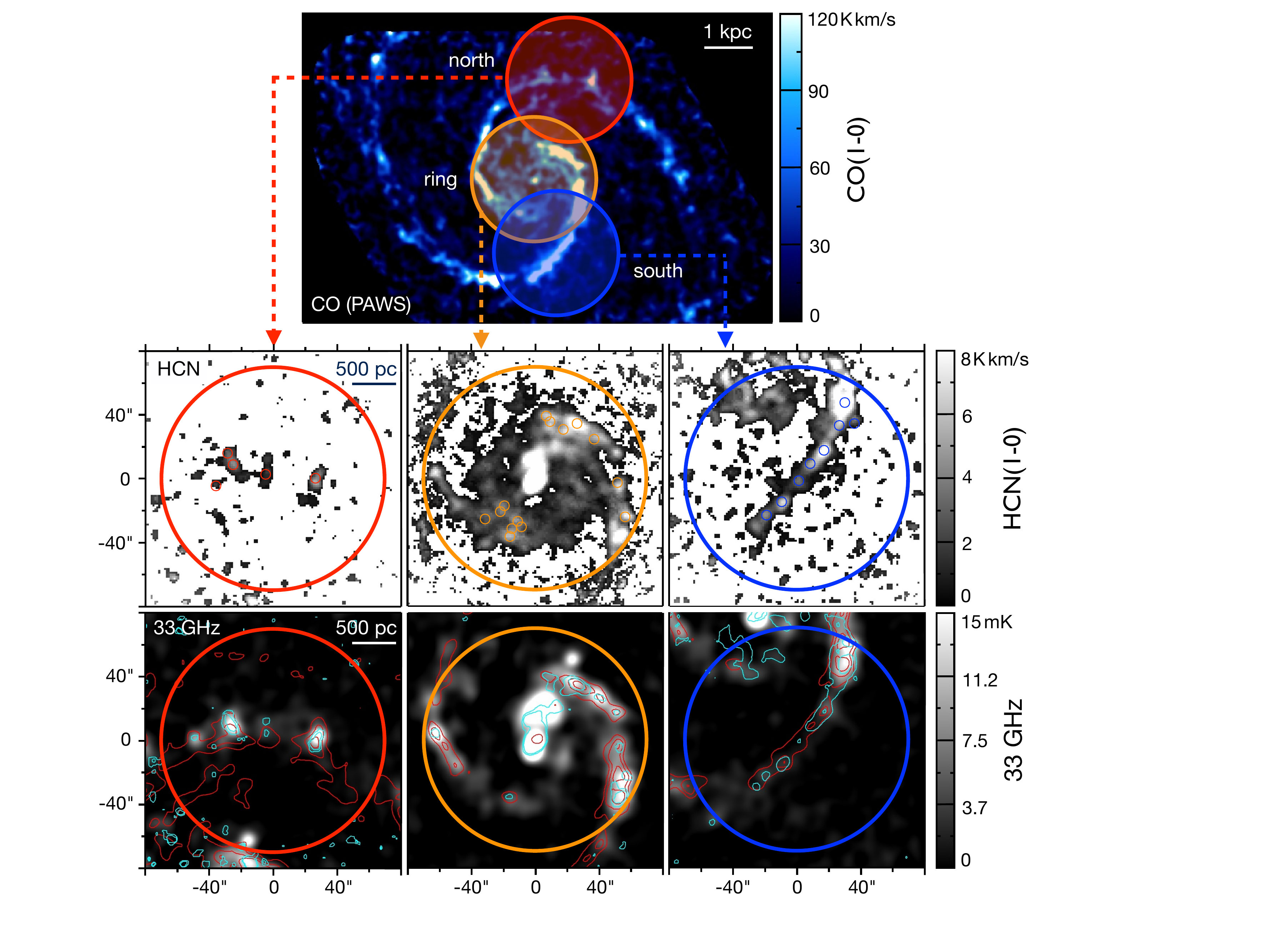}
\end{center}
\caption{\textit{Top panel:} PAWS map of the \mbox{CO(1-0)} line intensity in M51, highlighting the areas where we have mapped \mbox{HCN(1-0)} and which will be the focus of this paper. \textit{Middle panels:} Zoom on the targeted regions with \mbox{HCN(1-0)} intensity in grayscale, indicating the set of circular apertures centred on HCN peaks, with a diameter of $3''$ (FWHM of synthesised beam); the large, thick circles indicate the truncation radius of $R=35''$ (the same as the shaded areas in the top panel). \textit{Bottom panels:} background grayscale maps show the 33\,GHz radio continuum, tracing star formation for the same three regions, with contours showing CO emission (red) and HCN emission (cyan); CO contours correspond to [40, 80, 120]\,$\mathrm{K\,km\,s}^{-1}$ for the north, and [100, 150, 200]\,$\mathrm{K\,km\,s}^{-1}$ for the ring and the south, whereas HCN contours represent values of [1.5, 3]\,$\mathrm{K\,km\,s}^{-1}$ for the north, [5.5, 8]\,$\mathrm{K\,km\,s}^{-1}$ for the ring, and [3.5, 10]\,$\mathrm{K\,km\,s}^{-1}$ for the south.
}
\label{fig:regions}
\end{figure*}

To quantify spatial differences across the regions of interest, we define a number of circular apertures of diameter $3'' {\sim} 100$\,pc (the FWHM of our circular, synthesised beam). As we require that the apertures do not overlap significantly, these can be regarded as essentially independent, small-scale measurements across the disc of M51. The individual aperture measurements are provided in Table~\ref{table:results} and Table~\ref{table:results2}.

We are primarily interested in addressing the question of \textit{where does dense molecular gas form stars efficiently?} Therefore, we will mainly focus on apertures centred on local maxima in the map of \mbox{HCN(1-0)}. This choice is also motivated by the fact that we want to compare against the measurements from \citet{2017ApJ...836..101C}, who focussed on HCN-bright regions in the outer north spiral segment of M51. The local maxima are identified as the pixels of maximal intensity within isolated regions (``islands'') that arise on successively deeper cuts on the HCN intensity map (the moment-0 map described in Sect.~\ref{Sec:densedata}). This means that we start from a high threshold (a large multiple of the noise level), and iteratively reduce it until small, isolated regions appear in the thresholded map. The pixels of maximal intensity within the islands define the centres of the initial set of small circular apertures of fixed $3''$ diameter. We iteratively extend this procedure down to lower surface brightness cuts, and successively identify new islands whose area overlaps less than 20\% with any other island previously identified. We keep identifying new islands down to a threshold of ${\sim} 10\sigma$ for the ring pointing, ${\sim} 10\sigma$ for the southern area, and ${\sim} 6\sigma$ for the northern region (which is intrinsically fainter). With this method, we end up having a set of independent circular apertures of fixed $3''$ diameter that capture the strongest HCN emission peaks in each region (see Fig.~\ref{fig:regions}).

However, differences are expected if we place our apertures on CO or 33\,GHz peaks instead of HCN-bright regions \citep[e.g.][see Sect.~\ref{Sec:SFbreakdown}]{2010ApJ...722.1699S}. This is why in Table~\ref{table:tdep} we also quantify depletion times for apertures defined with the same procedure, but centred on CO or 33\,GHz peaks. Additionally, in the background of Figs.~\ref{fig:ICA-SFEdense}, \ref{fig:Sigma-SFE}, and \ref{fig:Fdense-SFR} we show as small open circles the measurements corresponding to all lines of sight with significant detections (down to $3\sigma$), using Nyquist-sampled, hexagonally-packed circular apertures across our three survey regions.

For all three maps (CO, HCN, and 33\,GHz) at matched $3''$ resolution, we multiply the measured fluxes by $\cos(i)=0.93$ so that surface densities refer to the actual plane of the galaxy, and are not relative to the projection on the sky. Thus, our apertures correspond to a deprojected area of 13,200\,pc$^2$. In the following plots, we identify any measurements below $3\sigma$ as upper limits (shown as open circles with downward arrows at the $3\sigma$ position).
In practice, for the apertures centred on HCN peaks, this only happens for some SFR measurements, while CO and HCN fluxes are always significant.

\section{Star formation efficiency and dense gas fraction} 
\label{Sec:SF-gas}

\subsection{HCN--TIR from Galactic to extragalactic scales}
\label{Sec:HCN-TIR}

\begin{figure*}[t]
\begin{center}
\includegraphics[trim=0 300 35  0, clip,width=1.0\textwidth]{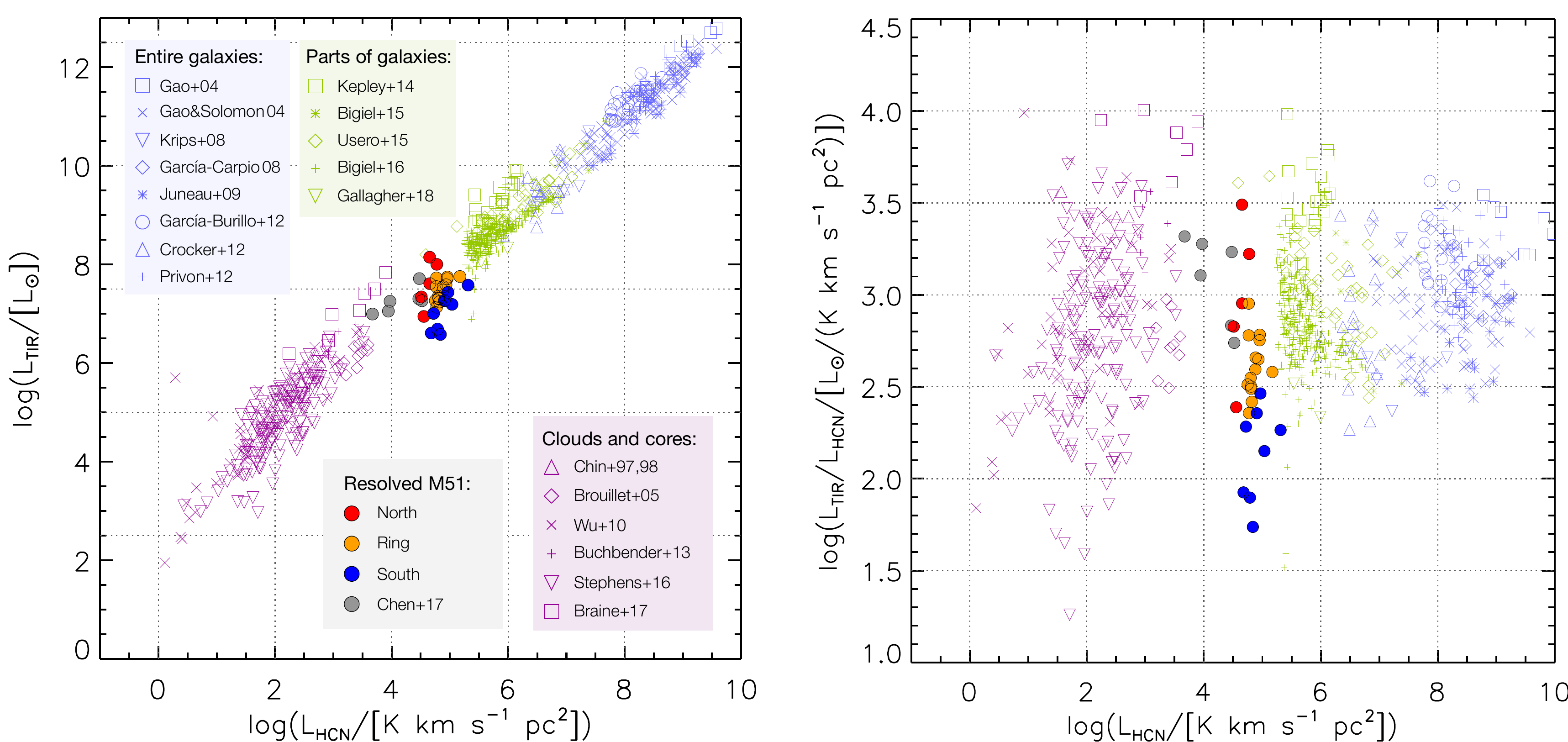}
\end{center}
\caption{\textit{Left panel:} TIR luminosity ($8{-}1000\,\mu$m), tracing the star formation rate, as a function of HCN luminosity, tracing dense molecular gas. \textit{Right panel:} TIR/HCN, tracing the dense gas star formation efficiency, as a function of HCN luminosity.
In both panels, we compare against measurements: (a) for whole galaxies \citep{2004ApJS..152...63G,2004ApJ...606..271G,2008A&A...479..703G,2008ApJ...677..262K,2009ApJ...707.1217J,2012MNRAS.421.1298C, 2012A&A...539A...8G, 2015ApJ...814...39P}; 
(b) kpc-size parts of galaxies \citep{2014ApJ...780L..13K,2015AJ....150..115U,2015ApJ...815..103B,2016ApJ...822L..26B,2018ApJ...858...90G}; 
(c) resolved clouds and cores \citep{1997A&A...317..548C,1998A&A...330..901C,2005A&A...429..153B,2010ApJS..188..313W, 2013A&A...549A..17B,2016ApJ...824...29S,2017A&A...597A..44B}. Our new datapoints in M51 at $3''$ resolution (${\sim} 100$\,pc) are displayed as filled circles, while the gray circles show similar-scale measurements in the outer spiral arm of M51 from \citet{2017ApJ...836..101C}.
}
\label{fig:TIR-HCN}
\end{figure*}

Since the seminal papers by \citet{2004ApJS..152...63G,2004ApJ...606..271G}, it has been recognised that a reasonably tight and linear correlation exists between the luminosity in the HCN emission line and the TIR of galaxies \citep[e.g.][]{2012A&A...539A...8G}. The linear correlation seems to persist down to sub-galactic scales of ${\sim} 10^{4.5}\,L_\odot$ in infrared luminosity, which corresponds to spatial scales of ${\sim} 1$\,pc, spanning almost 10\,dex in luminosity \citep{2005ApJ...635L.173W}.  We follow \citet{2013MNRAS.431.1956G} to convert 24\,$\mu$m luminosities to TIR ($8{-}1000\,\mu$m), and thus provide resolved aperture measurements for M51 in terms of HCN and TIR.

Figure~\ref{fig:TIR-HCN} (left panel) shows the 24\,$\mu$m-based TIR, a proxy for SFR, as a function of HCN molecular gas luminosity for our apertures in M51 centred on HCN peaks and a number of studies from the literature. We first show this plot to set the context of the discussion, because infrared luminosity has been the main tool in the field from star-forming cores to galaxies. However, we will shortly switch to 33\,GHz as the SFR tracer.
Our measurements at $3'' {\sim} 100$\,pc resolution fill the gap between the measurements for entire galaxies (light blue), kpc-size parts of galaxies (light green), and datapoints corresponding to resolved clouds and cores in the Milky Way or Local Group galaxies (purple). Our measurements cover a similar parameter space as  \citet{2017ApJ...836..101C} for a PdBI/NOEMA pointing at $150$\,pc resolution in the northern arm of M51. This apparent proportionality between star formation activity (tracked by TIR) and HCN luminosity (traditionally taken to trace dense molecular gas) has been used to argue in favour of simple density-threshold models, in which the intensity of star formation is set by the amount of gas above a certain density limit, which is converted into stars at a constant rate \citep{2004ApJ...606..271G,2005ApJ...635L.173W,2012ApJ...745..190L}.

However, as already highlighted in other studies \citep[e.g.][]{2015AJ....150..115U,2018ApJ...858...90G}, there is significant scatter in the HCN-TIR plane, and it encapsulates important physical differences. This is clearly demonstrated by the right panel of Fig.~\ref{fig:TIR-HCN}, which shows that the TIR/HCN ratio as a function of HCN luminosity exhibits major scatter.
\referee{It is worth noting that part of this scatter may arise from the different bands used to estimate TIR by different studies (24\,$\mu$m, 70\,$\mu$m, etc.). However,}
 in addition to significant scatter, our resolved $3''$ measurements in M51 also show systematic variations from region to region in the TIR/HCN ratio: there is a vertical offset between the environments that we target, the ``south'' clearly showing lower TIR/HCN. Even for comparable HCN luminosities, the offset between these datapoints spans almost 2\,dex, pointing to stark differences in the current star formation efficiency of the dense gas. 
We also see an increasing amount of lower outliers in the TIR/HCN ratio for the lowest $L_{\rm HCN}$; this effect was already discussed by \citet{2005ApJ...635L.173W}, who proposed that it could be reflecting a change in the sampling of dense gas masses relative to a basic unit of cluster formation. This increased scatter can also be partially explained as the result of sampling scales, because low HCN luminosities more likely correspond to individual regions which can reside in different evolutionary phases \citep{2014MNRAS.439.3239K}.

After this introductory section to set the context from Galactic to extragalactic scales, we will switch from TIR to our more direct approach to estimate SFRs using 33\,GHz continuum. It should not be understood that 24\,$\mu$m-based SFRs point to a qualitatively different picture from 33\,GHz; to a large extent, 24\,$\mu$m and 33\,GHz luminosities track each other (see Fig.\,\ref{fig:VLA-24um} and  Appendix\,\ref{sec:SFRcomparison}).
In the next sections, we will use the 33\,GHz tracer in conjunction with CO and HCN observations to explore trends in the star formation efficiency and dense gas fraction as a function of environment in M51.

\subsection{Spatial alignment of HCN, CO, and 33\,GHz peaks} 
\label{Sec:offsets}

\begin{figure}[t]
\begin{center}
\includegraphics[trim=140 160 170 180, clip,width=0.5\textwidth]{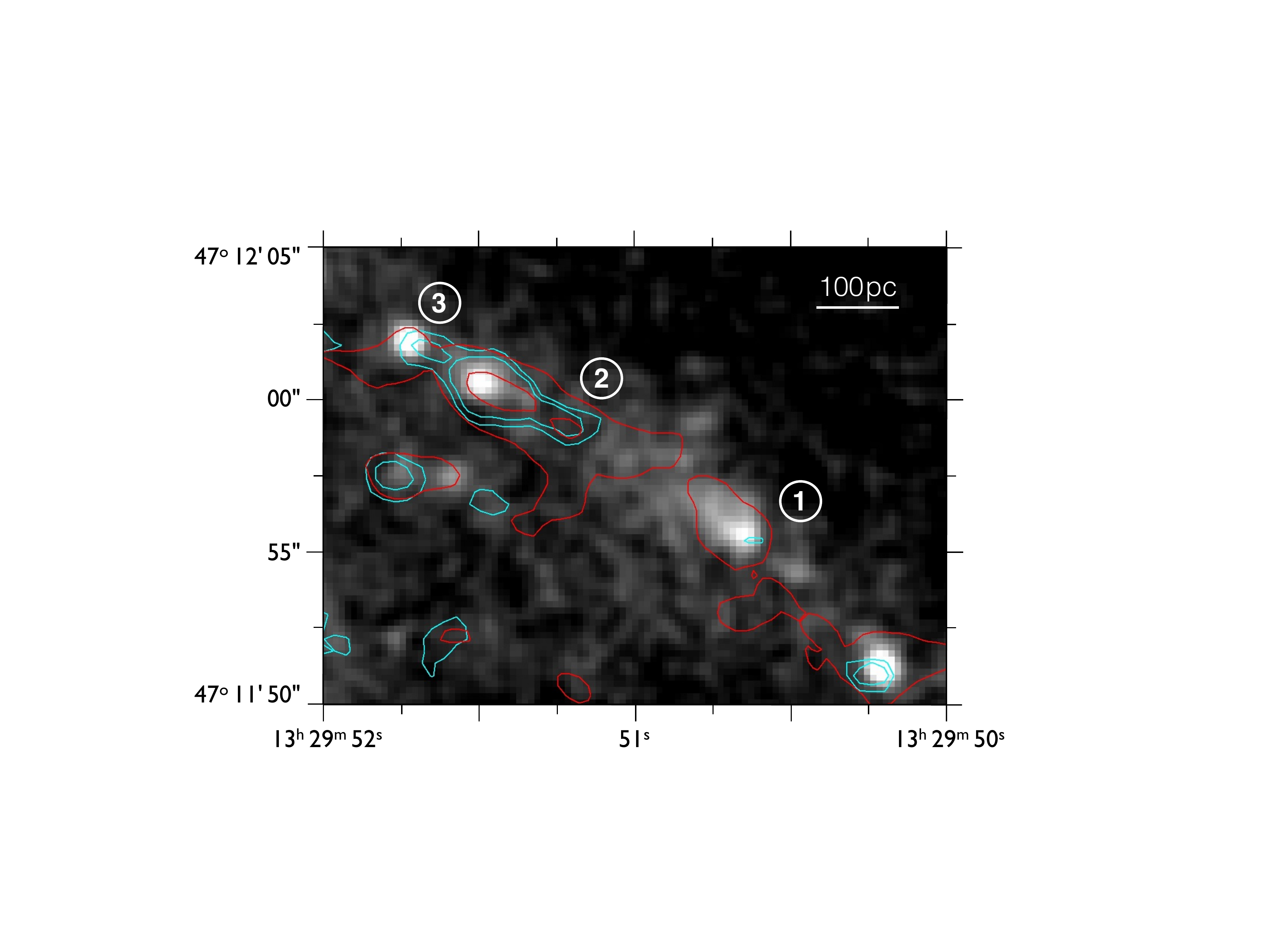}
\end{center}
\caption{Spatial offsets between gas and SFR tracers in the north-western part of the star-forming ring in M51. The background grayscale map is a ${\sim} 1''$ resolution version of the 33\,GHz continuum map. Cyan contours correspond to \mbox{HCN(1-0)} emission at ${\sim} 1.3''$ resolution (levels [6, 8]\,$\mathrm{K\,km\,s}^{-1}$), and red contours correspond to \mbox{CO(1-0)} emission at ${\sim} 1''$ resolution (levels [150, 350]\,$\mathrm{K\,km\,s}^{-1}$). 
}
\label{fig:offsets}
\end{figure}

The lower panel of Figure~\ref{fig:regions} shows the overlay of HCN and CO contours on the 33\,GHz map from the VLA at a matched resolution of $3''$. Even at this resolution, it becomes clear that 33\,GHz peaks do not always coincide with the highest intensity in HCN or CO. 
This is particularly clear at high resolution in the central region, where we have the highest quality data. 
For this experiment, we use the high-resolution HCN map (synthesised beam of $1.58'' \times 1.18''$), the 33\,GHz map imaged in natural weighting ($0.82'' \times 0.75''$), and the PAWS moment zero map at full resolution ($1.16'' \times 0.97''$).

Figure~\ref{fig:offsets} shows a blowup of the north-western part of the star-forming ring in M51, analogous to the middle panels of Fig.~\ref{fig:regions} but using the high-resolution (${\sim} 1''$) maps. In some areas, there seems to be an almost perfect spatial match between 33\,GHz peaks and local CO and HCN maxima (for example, the position marked \circled{1} in Fig.~\ref{fig:offsets}). However, in some other regions, we find CO and HCN peaks which do not seem to be associated with any significant 33\,GHz emission (position \circled{2} in Fig.~\ref{fig:offsets}). Finally, for some positions we find neighbouring peaks which are spatially offset (for instance, position \circled{3} in Fig.~\ref{fig:offsets} shows an offset of ${{\sim}}1'' \approx 40$~pc between the HCN and 33\,GHz peak). This agrees with the results from \citet{2010ApJ...722.1699S}, who showed that the CO / H$\alpha$ ratio (proportional to the molecular gas depletion time) in M33 diverges from the kpc-scale average when focussing on increasingly smaller apertures (resulting in shorter depletion times when the apertures are placed on H$\alpha$ rather than CO peaks).  Offsets between peaks of HCN emission and star formation traced by radio continuum have also been identified in other galaxies by \citet{2013ApJ...768...57P} and \citet{2015ApJ...813..118M}, and can be expected from the time evolution of the star formation process in individual clouds \citep{2014MNRAS.439.3239K,2018MNRAS.479.1866K}.

These spatial offsets of typically ${\lesssim} 1''$ in M51 suggest that the $3''$ apertures used for the measurements presented in the next sections encapsulate multiple regions which might be physically connected. 
We can further quantify the offsets between the brightest 33\,GHz and CO peaks at ${{\sim}}1''$ resolution not only in the nuclear ring, but across the whole PAWS field of view; we cannot do the same with HCN due to its lower resolution and the more limited field of view. The 33\,GHz peaks are easy to identify, as they mostly look like isolated point sources. For simplicity, we apply a high threshold on the VLA map ($25\,\mu$Jy/beam) that results in discrete islands of emission. Rather than defining CO peaks on the PAWS moment map, we use the catalogue of GMCs from \citet{2014ApJ...784....3C}, who used \texttt{CPROPS} \citep{2006PASP..118..590R} to measure the intensity-weighted mean CO position of the clouds at ${{\sim}}1''$ resolution.
We exclude from this analysis the AGN region (Sect.~\ref{Sec:SFRs}), and restrict the measurements to the PAWS field of view, which is slightly smaller than the extent of the VLA map.

This quantitative analysis confirms that the offsets between 33\,GHz and CO peaks are small, with a median of $1.2''$ (${\sim}50$\,pc). These offsets are simply the projected distance between the peak-intensity pixels of the 33\,GHz islands (on the highly thresholded VLA map) and the centroid of the nearest GMC in the PAWS catalogue, both at ${\sim} 1''$ resolution.
In fact, in 85\% of the cases, the 33\,GHz peaks have a GMC at a distance smaller than $2''$ ($<80$\,pc); in most of the 15\% remaining cases, the nearest GMC is actually closer to another 33\,GHz peak, which could mean that the original 33\,GHz peak has no detected GMC associated with it (although a given GMC could also be causally connected with several 33\,GHz peaks and vice versa). As stated previously, this is the natural consequence of time-evolution of individual star-forming regions \citep{2014MNRAS.439.3239K}, and we will  discuss it further in Sect.~\ref{Sec:SFbreakdown}.

\subsection{Star formation efficiency and dense gas fraction} 
\label{Sec:SFE}

Dense gas is a required step for star formation, but it is unclear exactly how its presence translates into a given amount of new stars. Does intense star formation across and within galaxies occur because more gas is dense, or because the available dense gas is more efficient at forming stars? In the first case, we would expect the star formation efficiency of the dense gas ($\mathrm{SFE_{dense}}$) to be roughly constant, as argued by density threshold models. Conversely, in turbulent models, the physical state of the dense gas is expected to play a key role in setting its ability to collapse and form new stars; in that case, $\mathrm{SFE_{dense}}$ is expected to change. Observations of the Galactic centre \citep[e.g.][]{2013MNRAS.429..987L,2014MNRAS.440.3370K} and of nearby galaxies at kpc-scales support the idea that $\mathrm{SFE_{dense}}$ is not constant \citep[e.g.][]{2015AJ....150..115U,2016ApJ...822L..26B,2018ApJ...858...90G}, and we aim to further test that with our data at higher resolution.

\referee{We remind the reader that we use our maps of \mbox{CO(1-0)} and \mbox{HCN(1-0)} to estimate the surface density of the bulk and dense molecular gas in the plane of the galaxy, $\Sigma_\mathrm{mol}$ and $\Sigma_\mathrm{dense}$, respectively (Sect.\,\ref{Sec:gassurfdens}). The star formation rate surface density, $\Sigma_\mathrm{SFR}$, is measured from free-free emission via 33\,GHz radio continuum (Sect.\,\ref{Sec:33GHz}).}
By ``star formation efficiency'' we refer to the star formation rate surface density divided by the (dense) molecular gas surface density ($\mathrm{SFE_{mol}} = \Sigma_\mathrm{SFR} / \Sigma_\mathrm{mol}$, $\mathrm{SFE_{dense}} = \Sigma_\mathrm{SFR} / \Sigma_\mathrm{dense}$). This efficiency is the inverse of the depletion time ($\tau_\mathrm{dep} = \Sigma_\mathrm{mol} / \Sigma_\mathrm{SFR}$), and we quote it in units of Myr$^{-1}$. We note that this is different from the dimensionless efficiencies that are often implemented in numerical simulations and that are more widespread in Galactic and theoretical work. Analogously, we define the ``dense gas fraction'' as the ratio of the dense molecular gas surface density inferred from HCN to the molecular gas surface density inferred from CO, i.e.\ $F_\mathrm{dense} = \Sigma_\mathrm{dense} / \Sigma_\mathrm{mol} \propto I_\mathrm{HCN} / I_\mathrm{CO}$. Therefore, by construction:

\begin{equation}
\mathrm{SFE_{mol}}=\mathrm{SFE_{dense}} \times F_\mathrm{{dense}},
\end{equation}

\noindent
which implies that the global star formation efficiency of the molecular gas is set by the product of the star formation efficiency of the dense gas and the dense gas fraction. In order to understand which of the two factors is more relevant ($\mathrm{SFE_{dense}}$ or $F_\mathrm{dense}$), and how each of them depends on environment, in the next sub-section we will analyse $\mathrm{SFE_{dense}}$ and $F_\mathrm{dense}$ as a function of stellar mass surface density and gas velocity dispersion in M51. As seen in previous studies, on kpc-scales $\mathrm{SFE_{dense}}$ anti-correlates while $F_\mathrm{dense}$ correlates with \referee{stellar mass surface density, which increases for decreasing galactocentric radius}; in addition, we concentrate on stellar mass surface density because it relates to midplane pressure and it facilitates comparison with previous studies.

Our measurements centred on HCN peaks at $100$\,pc resolution reveal relatively low star formation efficiencies, corresponding to long depletion times of ${\sim} 0.4 {-} 7$\,Gyr for the bulk molecular gas traced by CO (median value of 2.4\,Gyr; see Table~\ref{table:tdep}). This 2.4\,Gyr median value is slightly longer than the median  $\tau_\mathrm{dep}^\mathrm{mol} {\sim} 1.7$\,Gyr \citep{2012A&A...539A...8G},  $\tau_\mathrm{dep}^\mathrm{mol} {\sim} 2.2$\,Gyr \citep{2013AJ....146...19L},
$\tau_\mathrm{dep}^\mathrm{mol} {\sim} 1.1$\,Gyr \citep{2015AJ....150..115U}, or $\tau_\mathrm{dep}^\mathrm{mol} {\sim} 1.2$\,Gyr \citep{2018ApJ...858...90G} found by previous studies on ${\sim}$kpc scales (or a few hundred parsec). It is also considerably longer than the depletion time measured by \citet{2017ApJ...846...71L} over the whole M51 galaxy at 1.1\,kpc resolution (1.5\,Gyr with ${\sim}0.2$\,dex scatter), or the equivalent  measurement for the PAWS field of view, either at 1.1\,kpc resolution (1.7\,Gyr with ${\sim}0.1$\,dex scatter) or at 370\,pc resolution (very similar median, 1.6\,Gyr, but with ${\sim}0.3$\,dex scatter; \citealt{2017ApJ...846...71L}).

The star formation efficiencies of the dense gas traced by HCN span a similarly large dynamical range, corresponding to depletion times of ${\sim} 50 {-} 800$\,Myr (with median value of ${\approx}300$\,Myr for apertures centred on HCN peaks). 
This median dense gas depletion time of 300\,Myr is also longer than those from \citet{2015AJ....150..115U}, \citet{2012A&A...539A...8G}, and \citet{2018ApJ...858...90G} (median $\tau_\mathrm{dep}^\mathrm{dense} {\sim} 110$\,Myr, 140\,Myr, and 50\,Myr, respectively).

It is important to note that the literature studies at kpc-scales typically used TIR, H$\alpha$, and/or UV as the SFR tracer, whereas we are using 33\,GHz; therefore, systematic differences are likely (see Appendix~\ref{sec:SFRcomparison}).
Additionally, not surprisingly given the offsets that we have already examined in Sect.~\ref{Sec:offsets}, the absolute value of the depletion time depends strongly on the choice of apertures, as already demonstrated by \citet{2010ApJ...722.1699S} and quantified by \citet{2014MNRAS.439.3239K}; if we focus on apertures centred on the peaks of star formation, the depletion times become considerably shorter (median $\tau_\mathrm{dep}^\mathrm{mol} {\approx} 1.2$\,Gyr, $\tau_\mathrm{dep}^\mathrm{dense} {\approx} 130$\,Myr). Our results also demonstrate that focussing on HCN or CO peaks does not result in exactly the same depletion times, although both are longer than the depletion times measured for 33\,GHz peaks, as expected.

The dense gas fractions, on the other hand, are quite high. We measure dense gas fractions of $5{-}25\%$ in our $100$\,pc apertures (median ${\sim} 11{-}13\%$ for HCN and 33\,GHz peaks). These dense gas fractions are comparable to or higher than the kpc-scale median value of 12\% found by \citet{2012A&A...539A...8G}, 8\% from \citet{2015AJ....150..115U}, and 4\% from \citet{2018ApJ...858...90G}; however, if we bear in mind that \citet{2017ApJ...836..101C} found dense gas fractions of $2{-}5\%$ in the outer spiral arm of M51 at comparable resolution, we can conclude that our dense gas fractions are relatively high because we focussed on the central part of M51.

\referee{Table~\ref{table:rankcoeff1} summarises the strength  and statistical significance (through the Spearman rank coefficients) of the correlations that we will study in the subsequent sections. For our new observations, the correlation coefficients are reported for different sampling choices (all Nyquist-sampled detections, apertures centred on HCN peaks, CO peaks, and 33\,GHz peaks); we also report on the effect of simultaneously considering our measurements for HCN peaks and other samples from the literature at similar resolution (northern pointing in M51, M31, NGC\,3627). We compute rank coefficients excluding upper limits (values $<3\sigma$). We consider that a given correlation is statistically significant when the two-sided $p$-value is smaller than 5\% and, in Table~\ref{table:rankcoeff1}, we highlight these significant correlations in boldface.}

\begin{table*}[t!]
\begin{center}
\caption[h!]{Median depletion times of the bulk molecular gas, $\tau_\mathrm{dep}^\mathrm{mol}$ (Myr), of the dense gas, $\tau_\mathrm{dep}^\mathrm{dense}$ (Myr), and dense gas fraction, $F_\mathrm{dense}$, for $3''$ apertures centred on different peaks (HCN, CO, 33\,GHz, or unbiased Nyquist sampling) and different regions of M51.}
\begin{tabular}{lcccccccccccc}
\hline\hline
    & \multicolumn{4}{c}{$\tau_\mathrm{dep}^\mathrm{mol}$ (Myr)} & \multicolumn{4}{c}{$\tau_\mathrm{dep}^\mathrm{dense}$ (Myr)} & \multicolumn{4}{c}{$F_\mathrm{dense}$}  \\
  \cmidrule(lr){2-5} \cmidrule(lr){6-9} \cmidrule(lr){10-13} 
  & HCN  & CO  & 33\,GHz & unbiased & HCN  & CO  & 33\,GHz & unbiased & HCN  & CO  & 33\,GHz & unbiased  \\
  \hline  
M51 north        & 660   & 985 &  595 & 1179 &    105 & 88  & 57  & 73  &    10\% & 9\%  & 12\% &  10\% \\
M51 ring         & 2441 & 2535 & 1320 & 2677 &    286 & 155 & 148 & 210 &    14\% & 7\%  & 10\% &  11\%  \\
M51 south        & 5808 & 5808 & 2198 & 3635 &    582 & 593 & 383 & 468 &     9\% & 11\% & 17\% &  10\%  \\
 \hline
 M51 all         & 2441 & 2480 & 1244 & 2577 &    286 & 155 & 129 & 218 &    13\% & 8\%  & 11\% & 11\% \\
  \hline
\end{tabular}
\label{table:tdep}
\end{center}
\end{table*}

\subsubsection{Trends as a function of stellar mass surface density} 
\label{Sec:stellarmass}

\begin{figure*}[t]
\begin{center}
\includegraphics[trim=0 365 70 0, clip,width=1.0\textwidth]{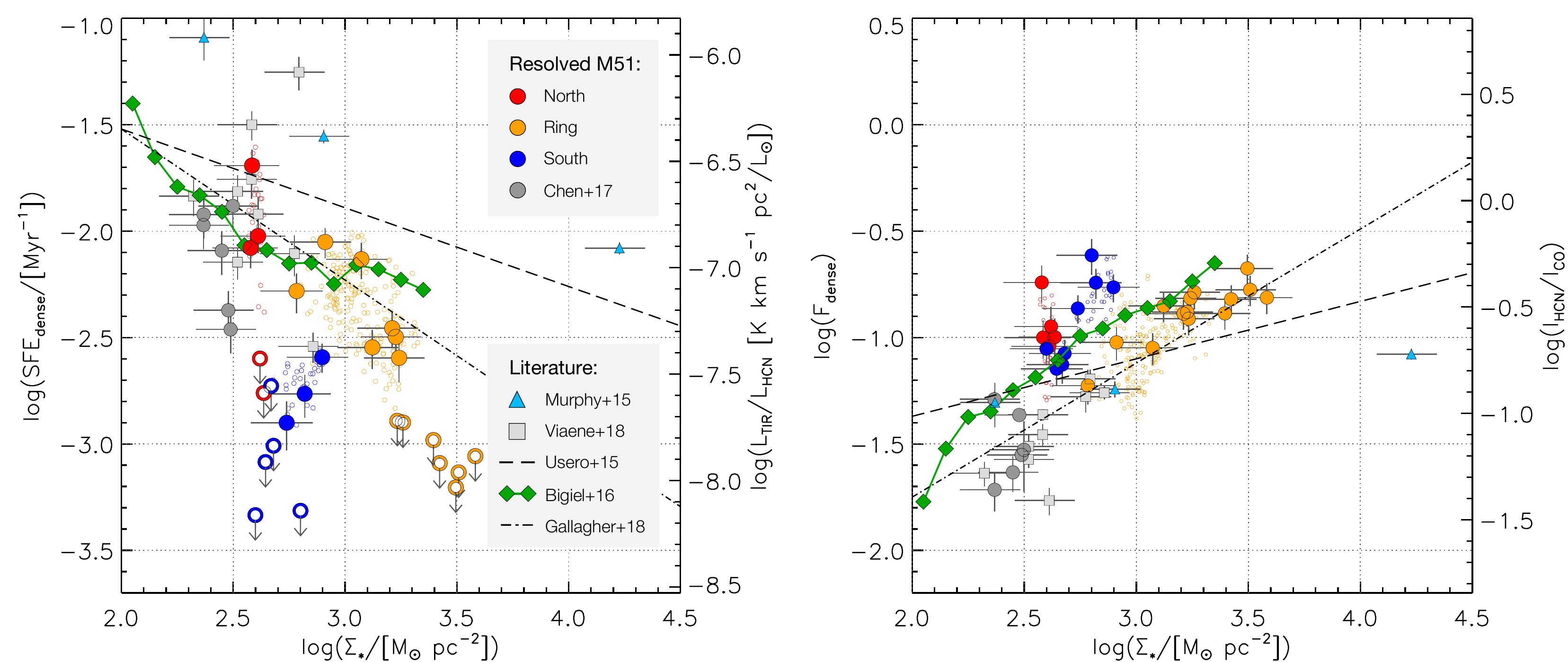}
\end{center}
\caption{Star formation efficiency of the dense gas ($\mathrm{SFE}_{\mathrm{dense}} = \Sigma_{\mathrm{SFR}} / \Sigma_{\mathrm{dense}}$, \textit{left panel}) and dense gas fraction ($F_{\mathrm{dense}} = \Sigma_{\mathrm{dense}} / \Sigma_{\mathrm{mol}}$, \textit{right panel}) as a function of local stellar mass surface density. The filled circles show results for $3''$ apertures (${\sim} 100$\,pc) centred on HCN peaks, while the small open circles in the background represent all detections across our fields of view (above $3\sigma$ simultaneously for $\mathrm{SFE}_{\mathrm{dense}}$, $F_{\mathrm{dense}}$, and $\Sigma_{\star}$). Open circles with a downward arrow indicate upper limits. The blue triangles and gray squares correspond to measurements at comparable scales for NGC\,3627 \citep{2015ApJ...813..118M} and M31 \citep[][and new SFRs described in the text]{2005A&A...429..153B}. The green diamonds connected by a solid line are the running medians from \citet{2016ApJ...822L..26B} from their single-dish map of the entire M51 disc ($28'' \sim 1$\,kpc resolution). The dashed black lines are the best-fit regressions from \citet{2015AJ....150..115U} for single-dish HCN observations of nearby galaxies at kpc-scale resolution. The dashed-dotted black lines are the fits to the ALMA observations presented in \citet{2018ApJ...858...90G} for four nearby galaxies, with a synthesised beam of a few hundred parsec sampled in radial bins.
}
\label{fig:ICA-SFEdense}
\end{figure*}

\begin{table*}[t!]
\begin{center}
\caption[h!]{Spearman rank correlation coefficients for several scaling relations.}
\begin{tabular}{lccccccc}
\hline\hline
  & $\mathrm{SFE_{dense}}$  vs. &  $F_\mathrm{dense}$  vs. & $\mathrm{SFE_{dense}}$  vs. &  $F_\mathrm{dense}$  vs. & $\mathrm{SFE_{mol}}$ vs &  $\Sigma_\mathrm{SFR}$  vs. &  $\Sigma_\mathrm{SFR}$  vs. \\
    & $ \Sigma_{\mathrm{\star}}$ &  $ \Sigma_{\mathrm{\star}}$ & $\sigma_{\mathrm{HCN}}$ & $\sigma_{\mathrm{HCN}}$ & $b_{\mathrm{CO}}$ & $F_\mathrm{dense}$ & $ \Sigma_{\mathrm{dense}}$   \\
  \hline  
M51: all detections	& $\boldsymbol{-0.24}$ (0.00) & 0.11 (0.07)       & $\boldsymbol{-0.42}$ (0.00) & \textbf{0.28} (0.00) &  0.06  (0.29) & $-0.04$  (0.42)	& \textbf{0.40}  (0.00) \\
$\bullet$ HCN peaks	& $-0.34$ (0.08) & \textbf{0.46} (0.02) & $\boldsymbol{-0.62}$ (0.03) & 0.15 (0.47) &  0.14  (0.65) & 0.04 (0.87) & $-0.06$  (0.84) \\
$\bullet$ CO peaks	&  0.28	 (0.30)	 & $-0.27$  (0.31) & $-0.52$  (0.07) & 0.34  (0.20) & 0.44  (0.15)	& 0.36 (0.23)	& 0.01  (0.97) \\
$\bullet$ 33\,GHz peaks & $-0.28$  (0.32) & 0.10  (0.72) 	& $-0.15$   (0.60)  & 0.50	 (0.06) & $-0.29$  (0.32)	& 0.16  (0.59) 	& 0.31  (0.27) \\
\hline
 M51 (incl. Chen+17)	& $\boldsymbol{-0.57}$ (0.00) & \textbf{0.70} (0.00) & ---            & --- 		&  ---  	& \textbf{0.63} (0.00) & \textbf{0.58} (0.00) \\
\hline
 M51, M31, NGC\,3627	& $\boldsymbol{-0.54}$ (0.00) & \textbf{0.71} (0.00) & $\boldsymbol{-0.53}$  (0.01) & \textbf{0.42} (0.01)	&  0.11  (0.61)	& \textbf{0.66}  (0.00) & \textbf{0.74} (0.00) \\
  \hline
\end{tabular}
\label{table:rankcoeff1}
\end{center}
\tablefoot{Spearman $\rho$ rank correlation coefficients \referee{(two-sided $p$-values)} for different apertures and regions in M51, as well as considering \referee{simultaneously our new measurements for HCN peaks in M51 and} comparable-scale measurements from the literature  (Fig.~\ref{fig:ICA-SFEdense}, Fig.~\ref{fig:Sigma-SFE}, Fig.~\ref{fig:Fdense-SFR}). We exclude upper limits from the analysis (values $<3\sigma$). \referee{Values in parenthesis indicate the $p$-value, and boldface} identifies the correlations that are statistically significant (two-sided $p$-value $< 5$\%).}
\end{table*}

On kpc-scales, both $\mathrm{SFE_{dense}}$ and $F_\mathrm{{dense}}$ have been shown to vary significantly among and within galaxies, clearly correlating with stellar mass surface density \citep{2015ApJ...810..140C,2015AJ....150..115U,2016ApJ...822L..26B,2018ApJ...858...90G}. We want to confirm if those correlations persist at 100\,pc scales, or if they break down when we approach the scales where gas and star formation peaks start to decouple from each other.

Fig.~\ref{fig:ICA-SFEdense} (left panel) shows the star formation efficiency associated with HCN (SFE$_{\mathrm{dense}}$) as a function of the local stellar mass surface density.
We overplot the results from \citet{2017ApJ...836..101C} for the outer spiral arm segment in the north of M51 (at a resolution of $4.9'' \times 3.7'' \sim 150-200$\,pc). We also overplot measurements at comparable physical scales (${\sim} 100$\,pc) from two other normal, star-forming galaxies: NGC\,3627 \citep{2015ApJ...813..118M} and M31 (molecular data from \citealt{2005A&A...429..153B} and SFRs from \citealt{2019ApJ...873....3T}). We deliberately avoid low-metallicity and starburst systems, which could introduce confusion in the trends due to chemistry. Given that systematic offsets exist among the different SFR tracers used by the studies from the literature, we rescale those SFRs by an empirically derived factor to enforce consistency (see Appendix~\ref{sec:SFRcomparison} for details). In NGC\,3627, for the nuclear pointing in \citet{2015ApJ...813..118M} the HCN line was not fully covered by their spectral set-up, so we use the information from \mbox{HCO$^+$(1-0)} instead in that particular case.

We find a decreasing trend between SFE$_{\mathrm{dense}}$ and stellar mass surface density, in the same sense as \citet{2015AJ....150..115U}, \citet{2016ApJ...822L..26B}, and \citet{2018ApJ...858...90G}; the linear regression to the data for a large sample of galaxies from Usero et al.\ is shown as a dashed line, the fit to Gallagher et al.\ as a dashed-dotted line,
and the running medians from Bigiel et al.\ as green diamonds. For \citet{2018ApJ...858...90G} this is a fit to their new ALMA observations of four nearby galaxies (NGC\,3551, NGC\,3627, NGC\,4254, and NGC\,4321), using H$\alpha + 24\,\mu$m as the SFR tracer (slope $-0.71$, intercept $-0.10$).
Even for the M51 data alone (including the points from \citealt{2017ApJ...836..101C}), we find a moderate degree of anti-correlation, and the anti-correlation is statistically significant ($\rho_{\mathrm{SFE_{dense}-\Sigma_\star}}=-0.57$, $p$-value\,$<0.1$\%); the Spearman rank correlation coefficient is slightly stronger than the one found by Usero et al.~on kpc-scales (they found $\rho_{\mathrm{SFE_{dense}-\Sigma_\star}} = -0.41$ to $-0.56$ depending on the SFR tracer), and slightly weaker than the one found by Gallagher et al.\ ($\rho_{\mathrm{SFE_{dense}-\Sigma_\star}} = -0.66$).
The degree of correlation slightly worsens when adding the datapoints for M31 and NGC\,3627, but it remains statistically meaningful and comparable to the results from Usero et al.~($\rho_{\mathrm{SFE_{dense}-\Sigma_\star}}=-0.54$, $p$-value\,$<0.1$\%). Significant galaxy-to-galaxy scatter has been observed before \citep[e.g.][Jim\'{e}nez-Donaire et al.~in prep.]{2015AJ....150..115U}. We have confirmed that these Spearman rank correlation coefficients are meaningful within $\Delta \rho \sim 0.1$ by running a Monte-Carlo simulation of 1000 trials where we randomly perturb the datapoints following a Gaussian distribution according to the uncertainties of the measurements. The rank coefficients are listed in Table~\ref{table:rankcoeff1}.

For a given stellar mass surface density, our datapoints span a wide range of SFE$_{\mathrm{dense}}$, exceeding 1\,dex for over most of the probed stellar mass surface densities. 
Even more importantly, SFE$_{\mathrm{dense}}$ shows clear \textit{systematic} variations from region to region: while the results for the northern spurs and those from \citet{2017ApJ...836..101C} tend to lie close to or above the running medians from \citet{2016ApJ...822L..26B}, the measurements for the star-forming ring and the southern arm cluster below the running medians.  Once again, the southern region clearly stands out as a strong lower outlier with respect to the overall trend. Since the ring and southern pointing share an overlap area (Fig.~\ref{fig:regions}), it is not surprising that there is some continuity between the measurements for the two regions, especially clear for the Nyquist-sampled apertures displayed in the background as small open circles.

In the right panel of Fig.~\ref{fig:ICA-SFEdense} we examine the ratio of the dense to bulk molecular gas surface densities traced by HCN and CO ($F_\mathrm{dense}$ being directly proportional to the observed ratio of their luminosities), as a function of the local stellar mass surface density. 
Again, the fit for \citet{2018ApJ...858...90G} is restricted to their four galaxies with new ALMA observations (slope 0.63, intercept $-3.01$).
 In agreement with previous studies, we find the dense gas fraction to increase with increasing stellar mass surface density. \referee{When we combine our measurements with those from \citet{2017ApJ...836..101C},} the correlation for M51 is strong and significant ($\rho_{\mathrm{F_{dense}-\Sigma_\star}}=0.70$, $p$-value\,$<0.1$\%). Interestingly, \citet{2015AJ....150..115U} also found a stronger correlation between $F_\mathrm{dense}$ and $\Sigma_\star$ ($\rho=0.67$) than between SFE$_{\mathrm{dense}}$ and $\Sigma_\star$ ($\rho \sim 0.5$), like we do here. This remains true if we include the datapoints from M31 and NGC\,3627 ($\rho_{\mathrm{F_{dense}-\Sigma_\star}}=0.71$, $p$-value\,$<0.1$\%). The degree of correlation between $F_\mathrm{dense}$ and $\Sigma_\star$ is comparable to the one found by \citet{2018ApJ...858...90G}, $\rho=0.67$.


Thus, summing up, stellar mass surface density seems to capture the mechanisms that set the dense gas fraction and the efficiency at which dense gas transforms into stars\referee{, probably because it is a tracer of mid-plane pressure}. However, rather than randomly scattering around a common relation, the measured star formation efficiencies in M51 seem to be systematically modulated by the details of the immediate galactic environment for a given stellar mass surface density (north, ring, south, outer arm). While on average we recover the trends observed at kpc-scales, when zooming in on 100\,pc scales, dynamical environment can lead to significant local differences. This suggests a direction that may help explain galaxy-to-galaxy scatter.
We note that the correlations with stellar mass surface density can, to a large extent, be equivalently expressed as a function of galactocentric radius, molecular gas surface density, or molecular-to-atomic gas fractions (since all of these observables closely correlate with local stellar mass surface density), as we discuss in Sect.~\ref{DiscussSFE}. 

\subsubsection{Trends as a function of velocity dispersion} 
\label{Sec:vdispersion}

\begin{figure*}[t]
\begin{center}
\includegraphics[trim=0 310 50 0, clip,width=1.0\textwidth]{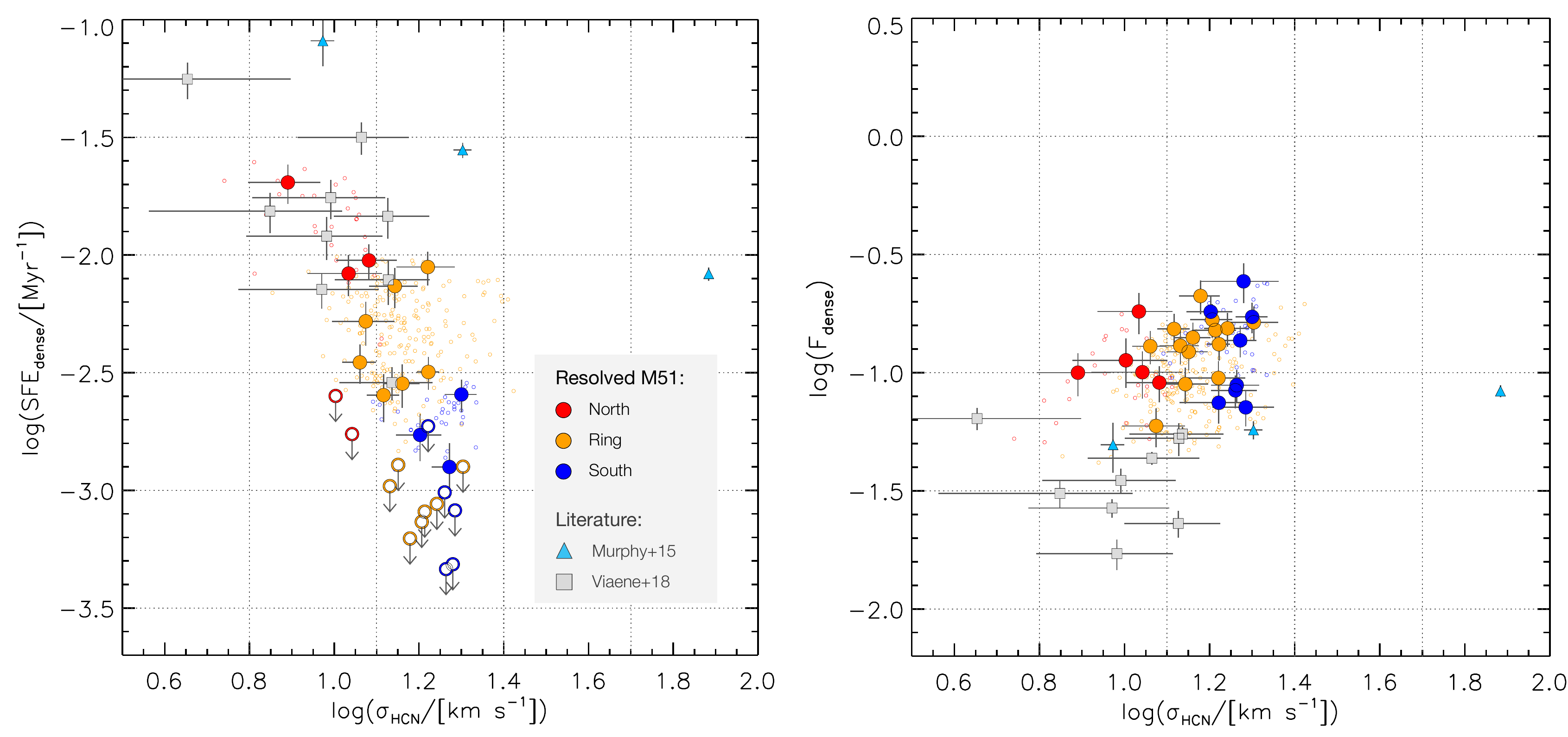}
\end{center}
\caption{Star formation efficiency of the dense gas ($\mathrm{SFE}_{\mathrm{dense}} = \Sigma_{\mathrm{SFR}} / \Sigma_{\mathrm{dense}}$, \textit{left panel}) and dense gas fraction ($F_{\mathrm{dense}} = \Sigma_{\mathrm{dense}} / \Sigma_{\mathrm{mol}}$, \textit{right panel}) as a function of the local velocity dispersion of HCN. The filled circles show results for $3''$ apertures (${\sim} 100$\,pc) centred on HCN peaks, while the small open circles in the background represent all detections across our fields of view (above $3\sigma$ simultaneously for $\mathrm{SFE}_{\mathrm{dense}}$, $F_{\mathrm{dense}}$, and $\sigma_{\rm HCN}$). The right panel shows more detections because the signal-to-noise ratio in CO and HCN is high compared to 33\,GHz (limiting the detections on the left panel). Open circles with a downward arrow indicate upper limits. The blue triangles and gray squares correspond to measurements at comparable scales for NGC\,3627 \citep{2015ApJ...813..118M} and M31 \citep[][and new SFRs described in the text]{2005A&A...429..153B}.
}
\label{fig:Sigma-SFE}
\end{figure*}

Here we study how the star formation efficiency behaves as a function of the velocity dispersion of the dense gas. Given the importance of turbulence in many current models of star formation 
\citep[e.g.][]{2004RvMP...76..125M, 2005ApJ...630..250K,2011ApJ...730...40P, 2012A&ARv..20...55H,2015MNRAS.450.4035F}, it is interesting to analyse the role of velocity dispersion in the cold molecular gas.
\citet{2017ApJ...846...71L} found a \textit{decreasing} trend between the star formation efficiency of the bulk molecular gas and CO velocity dispersion in M51. We investigate if this trend persists when looking at the star formation efficiency of the \textit{dense} gas traced here by HCN, or if it is instead driven by variations in the dense gas fraction. If the linewidths are mostly turbulent, the decreasing trend between star formation efficiency and CO velocity dispersion would suggest that star formation tends to be suppressed in highly turbulent gas.

Figure~\ref{fig:Sigma-SFE} (left panel) shows the star formation efficiency of the dense gas as a function of the velocity dispersion of HCN for our apertures \referee{centred on HCN peaks} and measurements from the literature on similar scales. These values quote the actual velocity dispersion, $\sigma$, in km/s, so that for a Gaussian profile the line width (FWHM) will be given by $2.35 \times \sigma$.
For M51, we only plot our new measurements, because \citet{2017ApJ...836..101C} did not provide velocity dispersions.

\referee{Our main result is that we find a significant anti-correlation between $\mathrm{SFE_{dense}}$ and $\sigma_\mathrm{HCN}$. This anti-correlation already holds for our new observations in M51 alone} ($\rho_{\mathrm{SFE_{dense}-\sigma_{HCN}}}=-0.62$, $p$-value\,$=2.5$\%), and it is stronger than the one between $\mathrm{SFE_{dense}}$ and $\Sigma_\star$. The anti-correlation becomes slightly weaker, \referee{but more significant,} when we simultaneously consider the measurements in M51, M31, and NGC\,3627 ($\rho_{\mathrm{SFE_{dense}-\sigma_{HCN}}}=-0.53$, $p$-value\,$<1$\%). \referee{The lower rank coefficient for the combined data} is largely driven by the offset datapoints from \citet{2015ApJ...813..118M}; considering M51 and M31 together, the anti-correlation is even stronger than for M51 alone ($\rho_{\mathrm{SFE_{dense}-\sigma_{HCN}}}=-0.71$ with $p<0.1$\%\referee{; not shown in Table~\ref{table:rankcoeff1}}).

\referee{The studies from the literature in M31 and NGC\,3627 measured velocity dispersions using HCN, $\sigma_\mathrm{HCN}$, and not CO. Thus, to consistently combine our results with the literature and expand the dynamic range, we also use HCN to measure velocity dispersions. Moreover, those studies targeted a few isolated sight-lines where HCN is bright enough to be detected; this is why we plot our measurements in M51 for apertures centred on HCN peaks, in order to keep the sampling analogous.}

\referee{However, if we ignore the datasets from the literature, we can test the effect of sampling and the choice of molecular gas tracer on our results in the inner part of M51. If we use $\sigma_\mathrm{CO}$ instead of $\sigma_\mathrm{HCN}$ but stick to the same apertures on HCN peaks (where CO is not necessarily brightest), the scatter increases and the anti-correlation between $\mathrm{SFE_{dense}}$ and $\sigma_\mathrm{CO}$ is no longer significant ($\rho_{\mathrm{SFE_{dense}-\sigma_{CO}}} =-0.10$, $p$-value\,$=75$\%; not shown in Table~\ref{table:rankcoeff1}). Conversely, if we measure the correlation between $\mathrm{SFE_{dense}}$ and $\sigma_\mathrm{CO}$ for apertures centred on CO peaks in M51, where the signal-to-noise ratio is higher, the correlation becomes significant again ($\rho_{\mathrm{SFE_{dense}-\sigma_{CO}}} =-0.71$, $p$-value\,$<1$\%; not listed in Table~\ref{table:rankcoeff1}).}

\referee{Our tests suggest that, when dealing with a limited dynamic range (like our three regions in M51), the choice of molecular gas tracer can affect the significance of the correlations. Through the choice of apertures, sampling also plays a role in setting the scatter and can therefore affect the derived correlation coefficients. Adding the measurements from the literature leads to more robust results, with higher statistical significance. We also confirmed that our} results are qualitatively robust against the method used to calculate the velocity dispersion; we checked that relying on second-order moment maps using the window method described in Sect.~\ref{Sec:densedata} results in comparable correlation coefficients. 

For completeness, the right panel of Fig.~\ref{fig:Sigma-SFE} shows the dense gas fraction as a function of the velocity dispersion \referee{traced by HCN}. 
\referee{Similarly to the trends with stellar mass surface density (Fig.~\ref{fig:ICA-SFEdense}), we find the reverse behaviour between $\mathrm{SFE_{dense}}$ and $F_\mathrm{dense}$ as a function of molecular gas velocity dispersion: $\mathrm{SFE_{dense}}$ decreases while $F_\mathrm{dense}$ increases for increasing $\sigma_\mathrm{HCN}$. However, in the case of $F_\mathrm{dense}$, the correlation with $\sigma_\mathrm{HCN}$ is only significant when we analyse simultaneously our measurements and the observations from the literature ($\rho_{F_\mathrm{{dense}}-\sigma_\mathrm{HCN}}=0.42$, $p$-value\,$<1$\%). If we only consider M51, the details of the scatter and the correlation coefficient depend again on the tracer used and the sampling choice.}

\citet{2017ApJ...846...71L} concluded that the boundedness parameter, $b=\Sigma_\mathrm{mol}/\sigma^2 \propto \alpha_{\rm vir}^{-1}$, probably reflects the dynamical state of molecular gas in M51 when measured on 40~pc scales. They also showed that it is a reasonably good predictor of $\mathrm{SFE_{mol}}$, in the sense that stronger self-gravity (higher $b$) leads to more efficient star formation.
We do not find a significant correlation with the boundedness parameter for our data ($\rho_{\mathrm{SFE_{mol}-b_{CO}}}=0.14$ with $p$-value\,$=65$\%); in principle, we cannot extend this analysis to the measurements from the literature, because they did not quantify the velocity dispersion using CO (in any case, if we use the velocity dispersion inferred from HCN to also include NGC\,3627 and M31, we would get $\rho_{\mathrm{SFE_{mol}-b_{CO}}}=0.11$ with $p$-value\,$=61$\%). 

One could think that the boundedness of the dense gas is physically more relevant than $b_\mathrm{CO}$ and might expect a relationship between $\mathrm{SFE_{dense}}$ and $b_\mathrm{HCN}$. However, we are clearly not resolving the scales where $\Sigma_\mathrm{dense}/\sigma_\mathrm{HCN}^2$ is representative of the boundedness of the dense gas. The physical expectation would be that the boundedness parameter derived from HCN is higher than the one derived from CO (i.e.\ dense gas is more bound); however, our data show the opposite, with $b_\mathrm{HCN}$ values which are lower than $b_\mathrm{CO}$.
This apparent contradiction can be explained by the insufficient resolution in the current extragalactic HCN observations to resolve individual bound dense gas units and, thus, the dense gas surface density is strongly beam-diluted (the observed $\Sigma_\mathrm{dense}$ is much lower when averaged inside the beam than in individual clumps). In addition, the fact that \referee{$\sigma_\mathrm{HCN}$ and $\sigma_\mathrm{CO}$ cover a similar range of values (despite individual differences)} suggests that $\sigma_\mathrm{HCN}$ is largely reflecting velocity dispersion among different clumps, and not the turbulent motions within a given clump; this can further lower the measured $b_\mathrm{HCN}$.

\referee{Summing up,} while the measurements from the southern spiral arm in M51 appeared as lower outliers to the $\mathrm{SFE_{dense}} - \Sigma_\star$ relation, they seem to follow the global decreasing trend in $\mathrm{SFE_{dense}} - \sigma_\mathrm{HCN}$. This might be indicative of turbulence and/or galactic dynamics playing a role in modulating how efficiently dense gas transforms into stars. We will discuss this further in Sect.\,\ref{Discussvdisp} and Sect.\,\ref{DiscussSouth}.

\subsection{Is the dense gas fraction a good predictor of star formation rate surface density at 100\,pc scales?} 
\label{Sec:Viaene}

\begin{figure*}[t]
\begin{center}
\includegraphics[trim=0 360 150 0, clip,width=1.0\textwidth]{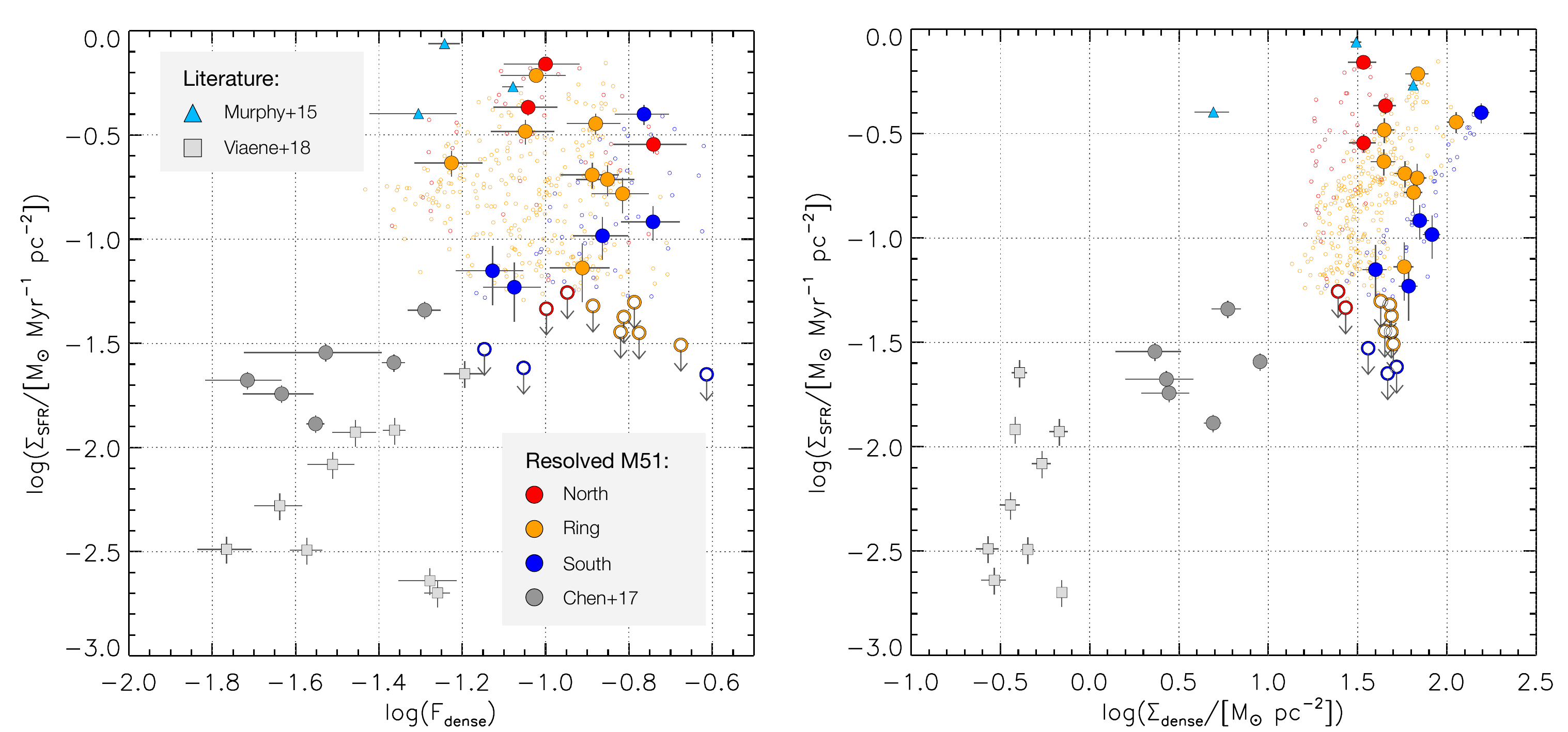}
\end{center}
\caption{Star formation rate surface density at ${\sim} 100$\,pc scales as a function of dense gas fraction ($F_{\mathrm{dense}} = \Sigma_{\mathrm{dense}} / \Sigma_{\mathrm{mol}}$, \textit{left panel}) and dense gas surface density ($\Sigma_{\mathrm{dense}}$, \textit{right}). The filled circles show results for $3''$ apertures (${\sim} 100$\,pc) centred on HCN peaks, while the small open circles in the background represent all detections across our fields of view (above $3\sigma$ simultaneously for $\Sigma_{\rm SFR}$, $F_{\mathrm{dense}}$, and $\Sigma_{\rm dense}$). Open circles with a downward arrow indicate upper limits. The blue triangles and gray squares correspond to measurements at comparable scales for NGC\,3627 \citep{2015ApJ...813..118M} and M31 \citep[][and new SFRs described in the text]{2005A&A...429..153B}.
}
\label{fig:Fdense-SFR}
\end{figure*}

Recently, \citet{2018MNRAS.475.5550V} showed that, for nine regions observed at 100\,pc resolution in M31, there is a good correlation between SFR and dense gas fraction (HCN/CO); the Spearman correlation coefficient was found to be $\rho=0.63$ (and as high as $\rho=0.98$ when removing a specific outlier). The authors used a suite of tracers to account for both obscured and unobscured star formation, and the dense gas masses were obtained from \mbox{HCN(1-0)} observations from the IRAM 30\,m telescope \citep{2005A&A...429..153B}. This result could suggest that the dense gas fraction might be a better predictor of star formation than the dense gas content. 

The observations that we present in this paper afford the possibility to extend this study to M51; our physical resolution of ${\sim} 100$\,pc is similar to that of \citet{2018MNRAS.475.5550V}. For the sake of the correlation examined by \citet{2018MNRAS.475.5550V} in M31, it was equivalent to consider integrated SFRs or SFR surface densities (because their apertures had a constant size). However, since M51 and M31 have very different inclinations, and the apertures from \citet{2017ApJ...836..101C} vary in size, we prefer to normalise the SFR by the area of the apertures (projected on the plane of each galaxy).
Fig.~\ref{fig:Fdense-SFR} (left panel) shows the SFR surface density, $\Sigma_\mathrm{SFR}$, against the dense gas fraction for $3''$ apertures (${\sim} 100$\,pc) in M51 centred on HCN peaks, analogous to Fig.~3 from \citet{2018MNRAS.475.5550V}. In addition to the  M31 datapoints (assuming an inclination of $i = 77^\circ$), we also plot the data from \citet{2015ApJ...813..118M} for NGC\,3627 (assuming $i = 62^\circ$). While \citet{2015ApJ...813..118M} obtained VLA (33\,GHz) and ALMA (HCN) maps at a resolution of ${\sim} 2''$, they carried out photometry after convolving the VLA and ALMA maps to the resolution of the CO(1-0) dataset from BIMA SONG ($7.3 \times 5.8'' \sim 300$pc at their assumed distance of $d = 9.38$\,Mpc). The right panel of Fig.~\ref{fig:Fdense-SFR} also shows $\Sigma_\mathrm{SFR}$ against the surface density of the dense gas, $\Sigma_\mathrm{dense}$, for the same datasets and apertures.

We find a significant correlation between $\Sigma_\mathrm{SFR}$ and $F_\mathrm{dense}$ when we simultaneously consider our new data and the measurements from \citet{2017ApJ...836..101C} in M51 ($\rho = 0.63$ with $p$-value\,$<1$\%). The correlation becomes slightly stronger when we expand the dynamic range by adding the datapoints for M31 and NGC\,3627 ($\rho = 0.66$ with $p$-value\,$<0.1$\%). We note that the degree of correlation worsens considerably for integrated SFRs instead of $\Sigma_\mathrm{SFR}$ ($\rho = 0.33$), mostly as a result of the different sizes of the apertures used by \citet{2017ApJ...836..101C}. We remind the reader that we use the SFRs from \citet{2019ApJ...873....3T}, but the global conclusions do not qualitatively change if we use the SFRs from Viaene et al.~instead. The combined data in the right panel of Fig.~\ref{fig:Fdense-SFR} show that the correlation between $\Sigma_\mathrm{SFR}$ and $\Sigma_\mathrm{dense}$ is even stronger ($\rho = 0.74$) than the correlation between $\Sigma_\mathrm{SFR}$ and $F_\mathrm{dense}$.
Therefore, judging by the data available to us, the dense gas fraction does not seem to be a better predictor of the SFR surface density at 100\,pc scales than the dense gas surface density.


\section{Limitations and caveats} 
\label{Sec:caveats}

\subsection{33\,GHz continuum as a tracer of star formation} 
\label{Sec:usefultracer}

Free-free radio emission has been proposed as a ``gold standard'' to trace star formation activity in galaxies because its flux is proportional to the production of ionising photons in newborn stars, without the necessity to resort to indirect, empirical calibrations. Moreover, free-free emission is not subject to extinction problems which complicate the estimation of SFRs at the ultraviolet and optical wavelengths. 
We have estimated how the thermal fraction for 33\,GHz continuum emission varies from region to region, and used the thermal free-free emission to map star formation in the spiral galaxy M51.
It is important to emphasise that free-free emission is a tracer of high-mass star formation, in the sense that it is only sensitive to stars which are capable of ionising the surrounding gas to produce an \ion{H}{II} region; 
this is why only for a fully sampled IMF will the ionising photon production rate be proportional to the rate of recent  star formation (over timescales of ${\lesssim} 10$\,Myr), a limitation that also applies to H$\alpha$ as a SFR tracer. The conversion between 33\,GHz flux and ionising photon luminosity in \citet{2011ApJ...737...67M} relies on the assumption of Starburst99 stellar population models \citep{1999ApJS..123....3L} with a Kroupa IMF \citep{2001MNRAS.322..231K}. It also relies on the analysis in \citet{1968ApJ...154..391R}, which only accounted for ionised hydrogen and not for ionised helium. A different choice of IMF, stellar population models, or the inclusion of helium would introduce a systematic offset in the inferred SFRs, but it should not vary from region to region and therefore it will not affect the trends that we find \citep[see also][]{2007ApJ...666..870C}.

On the other hand, there are some caveats associated with the use of this wavelength range to trace star formation. Firstly, not all of the continuum at 33\,GHz is arising from free-free emission; synchrotron emission can also contribute a non-negligible fraction of the flux. This is especially true around AGN, which is why we have excluded from our analysis the inner part of M51, surrounding its Seyfert-2 nucleus. However, excluding active nuclei, 33\,GHz is expected to be dominated by thermal emission; for example, \citet{2011ApJ...737...67M} found an average 87\% thermal fraction for a sample of nearby galaxies \citep[see also][]{2010ApJ...709L.108M}. In any case, for M51 we have access to radio observations in other bands \citep[e.g.\ 20\,cm, 6\,cm, and 3.6\,cm;][]{2011AJ....141...41D} and we have used them to explicitly estimate the thermal fraction for each of our apertures through the spectral index. Our estimate relies on the assumption of a fixed non-thermal spectral index ($\alpha^\mathrm{NT} = 0.85$), but this seems to be quite stable for typical star-forming galaxies \citep{1997A&A...322...19N,2011ApJ...737...67M}. 
The typical uncertainty in the estimation of the thermal fraction is ${\sim} 10$\% (the individual uncertainties are quoted in Table~\ref{table:results}). Additionally, one needs to assume an electron temperature when converting thermal 33\,GHz fluxes into SFRs (Eq.~(\ref{eq:conversionSFR})), but this is well constrained in M51, and it introduces an uncertainty of only ${\lesssim} 5\%$ on the derived SFRs (Sect.~\ref{Sec:SFRs}).

We do not expect substantial flux filtering associated with our (very compact) configuration of the VLA interferometer, which is sensitive to spatial scales up to $44'' \approx 1.6$\,kpc. \citet{2018ApJS..234...24M} showed that there is not a large amount of missing flux comparing 112 star-forming regions targeted both with the VLA and the GBT, which, as a single dish, should recover all the flux; this study includes some pointings in M51. In any case, if flux filtering is taking place due to missing short spacings, this can be expected to be mostly diffuse 33\,GHz flux, which is more likely associated with synchrotron emission than with free-free emission arising from compact \ion{H}{II} regions.

A potential caveat when estimating SFRs from free-free emission is that part of the Lyman-continuum photons produced by massive stars may be absorbed inside \ion{H}{II} regions before they reach a hydrogen atom. This caveat also applies to recombination-line tracers such as H$\alpha$ or Pa\,$\alpha$, because the UV photons absorbed by dust would not be available to ionise hydrogen. Alternatively, some UV photons may leak out of a given \ion{H}{II} region to the diffuse ISM before ionising hydrogen. Some Galactic and Local Group work suggests that up to half of the Lyman continuum photons might not contribute to hydrogen ionisation within \ion{H}{II} regions \citep{2001AJ....122.1788I,2001ApJ...555..613I,2018ApJ...864..136B}. However, a number of extragalactic papers reassuringly found good agreement between the Lyman-continuum luminosities (or SFRs) inferred from thermal fluxes and infrared measurements on kpc scales \citep{1994ApJ...421..122T,2011ApJ...737...67M,2012ApJ...761...97M}. This would imply that only a minor fraction of ionising photons is absorbed inside \ion{H}{II} regions. 
It is unclear if the discrepancy between the Local Group and nearby spiral galaxies is due to observational biases (e.g.\ complications in the Galactic foreground subtraction), due to the vastly different scales probed, or other factors. In any case, this issue could be far more problematic for compact starbursts hosted by (U)LIRGs \citep{2003ApJ...583..727D}.

An additional complicating factor associated with 33\,GHz emission can be the presence of ``anomalous dust emission.'' \citet{2010ApJ...709L.108M} targeted a few regions in NGC\,6946 and showed that, while 33\,GHz continuum is typically dominated by free-free, 33\,GHz can occasionally show significant excess flux which can be attributed to dipole emission from rapidly rotating ultra-small grains. \citet{2010ApJ...709L.108M} found a higher flux density in the 33\,GHz band than in the 8.4\,GHz band for the case of anomalous dust emission (the extranuclear region 4 in NGC\,6946). However, for our apertures in M51, we do not find even a single occurrence in which the 33\,GHz flux density is higher than the flux density in the neighbouring 8.4\,GHz band. This is reassuring, and although it does not fully rule out the presence of some anomalous dust emission in our measurements, it suggests that the contribution in the regions that we target must be fairly limited, if any.


The fact that 24\,$\mu$m (especially when translated to TIR) results in considerably higher SFRs than 33\,GHz (see Appendix~\ref{sec:SFRcomparison}) could be partially due to dust heated by old stellar populations \citep{2012AJ....144....3L,2017A&A...599A..64V}.
This old stellar contribution to the infrared has been recently highlighted, precisely in the case of M51, by \citet{2017ApJ...851...10E}.
\citet{2016A&A...591A...6B}  also find that hybrid UV$+$IR SFR estimators (e.g.\ UV$+$70\,$\mu$m) strongly depend on stellar mass surface density. This could lead to a particularly large discrepancy in the central area of M51 that we have targeted, where high stellar mass surface densities are reached. \citet{2017ApJ...846...71L} also compared several SFR tracers in M51, concluding that TIR may somewhat overestimate SFRs.  On the other hand, the \citet{2007ApJ...666..870C} prescription for 24\,$\mu$m was calibrated on slightly larger scales (${\sim}$500\,pc) than our measurements, averaging out finer details, and, strictly speaking, it is not directly applicable to our 100\,pc apertures.
In any case, given the caveats that we have discussed before, 33\,GHz could have a tendency to underestimate SFRs (e.g.\ flux filtering, escape of ionising photons), while 24\,$\mu$m probably tends to overestimate them; the combined effect could explain the measured offset in SFRs.

\subsection{Does \mbox{HCN(1-0)} trace dense molecular gas in M51?} 
\label{Sec:HCNproblems}

To first order approximation, the high critical density of HCN \citep[$n_\mathrm{crit}^\mathrm{HCN} {\sim} 10^6$\,cm$^{-3}$, compared to $n_\mathrm{crit}^\mathrm{CO} {\sim} 10^3$\,cm$^{-3}$;][]{2015AJ....150..115U,2017A&A...599A..98P} would suggest that HCN emission arises from molecular gas at much higher densities than the regions traced by CO. 
However, it has often been pointed out that HCN can be excited at lower densities due to effects of opacity and ISM structure: at least as low as $n_\mathrm{eff}^\mathrm{HCN} {\sim} 10^4-10^5$\,cm$^{-3}$ \citep{2015AJ....150..115U,2017MNRAS.466...49J,2017ApJ...835..217L}; this effect is also well known from Galactic studies \citep[see e.g.][]{2015PASP..127..299S}. 

From a Galactic observational perspective, some authors have further questioned the idea that emission from molecules such as HCN traces gas at such high densities \citep{2017A&A...605L...5K,2017A&A...599A..98P,2017A&A...604A..74S}. Specifically, the kinetic temperature dependence of the \mbox{HCN(1-0)} line excitation can make it brighter next to \ion{H}{II} regions as a result of increased far-UV (FUV) illumination \citep{2017A&A...599A..98P}. In that case, there might be a degeneracy between FUV illumination and high density when it comes to exciting \mbox{HCN(1-0)}, as \citet{2018A&A...610A..12B} argue based on multi-line observations of the Orion\,B molecular cloud, and using a machine-learning clustering method to differentiate between physical and chemical regimes. From the Galactic point of view, N$_2$H$^+$ has been argued to be an excellent tracer of the densest gas  \citep[e.g.][]{2017A&A...599A..98P}, but those observations would be extremely time-consuming for external galaxies and are currently not available for M51.

In any case, these studies are based on specific star-forming regions in the Milky Way (e.g.\ Orion~A, Orion~B), and it is not fully clear how their conclusions would extrapolate to extragalactic studies, which often average over many such regions within a single beam.  Specifically, there is mounting evidence that CO emission in nearby galaxies is probing a combination of moderately dense gas associated with GMCs and a significant fraction of diffuse molecular gas, arising from presumably less dense, spatially extended structures \citep[e.g.][]{2013ApJ...779...43P,2015AJ....149...76C,2016AJ....151...34C}. If the excitation conditions prevalent in this diffuse gaseous component do not result in significant HCN emission, this might contribute to explaining why HCN can be a better tracer of dense gas than CO at the range of scales typically sampled by extragalactic studies (because CO at coarse physical resolutions can be biased by diffuse gas). 

In the extragalactic context, it is unavoidable to average over regions of different density inside one synthesised beam.
\citet{2017ApJ...835..217L} recently showed that, for a log-normal gas density distribution, most HCN emission can arise from only moderately dense gas when the density distribution peaks below the critical density. But even in those cases the HCN/CO ratio will also be low, implying that the observed flux ratios are likely more robust than estimates of dense molecular gas masses (or surface densities). Adding a power-law tail to the density distribution tends to stabilise the median density for emission, suggesting that in those cases HCN emission arises mostly from gas at ${\sim} 10^4-10^5$\,cm$^{-3}$. 
 \citet{2018ApJ...858...90G} provided additional observational support to this idea with ALMA data for four nearby galaxies. They found high HCN/CO ratios in regions of high CO surface brightness (as expected if high dense gas fractions occur in the regions of high mean gas density); they also found that HCN/CO correlates well with HCO$^+$/CO (whereas if abundance variations are unrelated to density, HCO$^+$ might show significant departures from HCN). The correlation between the HCN/CO line ratio and the cloud-scale surface density was systematically quantified for five nearby galaxies in \citet{2018ApJ...868L..38G}, further supporting the idea that the ratio HCN/CO is sensitive to gas density. 
 In any case, more work is needed to better understand how all these factors interact and set the values of $\alpha_\mathrm{CO}$ and $\alpha_\mathrm{HCN}$ in detail within galaxies.

\section{Discussion} 
\label{Sec:discussion}


\subsection{Trends with stellar mass surface density} 
\label{DiscussSFE}

At kpc-scales, both the star formation efficiency of the dense molecular gas ($\mathrm{SFE_{dense}}$) and the dense gas fraction ($F_\mathrm{{dense}}$) have been shown to vary significantly among and within galaxies; their product sets the overall star formation efficiency, which we define as the inverse of the molecular gas depletion time. Our results confirm strong variations in $\mathrm{SFE_{dense}}$ and $F_\mathrm{dense}$ at ${\sim} 100$\,pc scales in M51; the variation is particularly extreme in $\mathrm{SFE_{dense}}$, which spans almost 2\,dex. 
This strong variation in $\mathrm{SFE_{dense}}$ is hard to reconcile with simple density threshold models, where the amount of gas above a certain critical density sets the rate at which new stars form \citep{2004ApJ...606..271G,2005ApJ...635L.173W,2012ApJ...745..190L}. Previous work carried out at lower spatial resolution also found discrepancies with density threshold models \citep[e.g.][]{2015AJ....150..115U,2016ApJ...822L..26B,2018ApJ...858...90G}. 


Moreover, $\mathrm{SFE_{dense}}$ and $F_\mathrm{{dense}}$ have been found to depend on environment, both among and within galaxies, closely correlating with stellar mass surface density on kpc-scales \citep[e.g.][]{2015AJ....150..115U,2015ApJ...810..140C,2016ApJ...822L..26B,2018ApJ...858...90G}. These trends can be recast as a function of galactocentric radius, molecular-to-atomic gas fraction, molecular gas surface density, or dynamical equilibrium pressure \citep{1989ApJ...338..178E,1997ApJ...478..162H,2010ApJ...721..975O,2011ApJ...743...25K}. For example, we find a similarly strong correlation between $\mathrm{SFE_{dense}}$ or $F_\mathrm{{dense}}$ and $\mathrm{\Sigma_{mol}}$ as we did with $\Sigma_{\star}$ ($\rho = - 0.57$ and 0.51, respectively, when considering simultaneously the data for M51, M31, and NGC\,3627; $p$-value $< 0.1$\%).
The dependence of $\mathrm{SFE_{dense}}$ on environment goes in the same direction as results from the Milky Way, where dense gas in the Central Molecular Zone has been shown to be less efficient at forming stars than in the solar neighbourhood \citep[e.g.][]{2013MNRAS.429..987L,2013ApJ...765L..35K,2014MNRAS.440.3370K,2014ApJ...795L..25R}.

It was unclear if such trends would be recovered at 100\,pc scales due to the increased stochasticity in the sampling of the star formation process (see Sect.~\ref{Sec:SFbreakdown} below). However, combining our new measurements for three regions of M51 with the results of \citet{2017ApJ...836..101C} for an outer spiral arm segment of M51, we recover statistically significant correlations which are in qualitative agreement with those previously found: when we move to regions of higher stellar mass surface density, $F_\mathrm{{dense}}$ increases while $\mathrm{SFE_{dense}}$ decreases. 
In fact, the rank correlation coefficients that we find are comparable to those found at kpc-scales by \citet{2015AJ....150..115U}. These correlations remain true if we include measurements from other galaxies on similar scales (${\sim}100{-}300$\,pc; M31, NGC\,3627).
In any case, differences with respect to Usero et al.~and similar studies could be expected not only as a result of the different scales that we probe, but also due to the different SFR tracers used (see Sect.~\ref{Sec:usefultracer}), and due to intrinsic galaxy-to-galaxy variations. It is remarkable that we find such similar correlations when examined at ${\sim}100$\,pc instead of kpc scales.

\subsection{Trends with velocity dispersion} 
\label{Discussvdisp}

\citet{2015AJ....150..115U} quantitatively demonstrated that the dependence of $\mathrm{SFE_{dense}}$ and $F_\mathrm{{dense}}$ on environment is compatible with models of turbulent star formation \citep[e.g.][]{2005ApJ...630..250K,2007ApJ...669..289K,2012ApJ...761..156F}. As opposed to models advocating for a simple density threshold for star formation \citep[e.g.][]{2004ApJ...606..271G,2005ApJ...635L.173W,2012ApJ...745..190L}, turbulent models claim that the physical state of the cloud, and not only its density, affects its ability to collapse and form stars (for example, through the turbulent Mach number, particularly if turbulence is primarily compressive; see \citealt{2017A&A...599A..99O}). 
We observe a significant anti-correlation between the star formation efficiency of the dense gas and the linewidth of HCN (Fig.~\ref{fig:Sigma-SFE} and Table~\ref{table:rankcoeff1}). This is true for our new measurements in M51 ($\rho_{\mathrm{SFE_{dense}-\sigma_{HCN}}}=-0.62$), and it remains true when we include the values for other galaxies from the literature ($\rho_{\mathrm{SFE_{dense}-\sigma_{HCN}}}=-0.53$). The trend goes in the same direction as that already pointed out by \citet{2015ApJ...813..118M} based on HCN emission in NGC\,3627 (with three datapoints at ${\sim}300$\,pc resolution), although there is a vertical offset between the trend in NGC\,3627 and that in M51. Our measurements extend to the dense gas phase the result from \citet{2017ApJ...846...71L}, who already identified in M51 a correlation between the velocity dispersion and star formation efficiency of the bulk molecular gas traced by CO (using the full PAWS resolution, 40\,pc), in the sense that larger linewidths are associated with lower star formation efficiencies.

In agreement with our observations, the SFE$_\mathrm{dense}$ predicted by the model of Meidt et al.\ (in prep.) decreases with increasing $\sigma$. This model reflects the competition between gas self-gravity and motions in the galactic potential, which dominate the gas kinematics on large scales.  The degree to which the dense gas can become self-gravitating (and star-forming) at fixed $\sigma$ depends on the local gas surface density and the dense gas fraction in the model. In this context, the lower dense gas fraction in NGC\,3627 ($F_\mathrm{{dense}} \sim 6\%$) could explain the vertical offset of the \citet{2015ApJ...813..118M} datapoints relative to our M51 measurements. 

\citet{2017ApJ...840...48P} presented a revised model based on \citet{2011ApJ...730...40P}, which predicts a decreasing trend between the SFE per free-fall time and the virial parameter. For individual clouds, ${\rm SFE_{ff}} = 0.4\,\exp(-1.6 \alpha_{\rm vir}^{1/2})$ was found to successfully describe the properties of star-forming regions in a high-resolution numerical simulation of turbulent gas (albeit with large scatter), and it was found to agree well with the predictions from the model. At fixed molecular gas surface density, this model would imply an exponentially decreasing SFE as a function of velocity dispersion, which might be in agreement with our observations; other numerical simulations and models have also found a decreasing trend between SFE and $\sigma$ \citep[e.g.][]{2015MNRAS.451.3679B,2018ApJ...863..118B}.

While higher turbulence should result in increased linewidths, the inverse is not true in general, as unresolved streaming motions can also contribute to the 
observed linewidths. 
This is especially true at low or intermediate resolutions, 
as discussed by \citet{2017ApJ...846...71L} for M51. In any case, 
if linewidths are primarily reflecting turbulent motions in the gas, 
the anti-correlation between $\mathrm{SFE_{dense}}$ and $\sigma$
might be challenging for some turbulent models, which tend to predict an increase in the star formation efficiency as a function of turbulent Mach number \citep[e.g.][]{2005ApJ...630..250K,2011ApJ...743L..29H,2013ApJ...763...51F}.

\citet{2018ApJ...860..172S} recently demonstrated that a strong correlation exists between linewidth and molecular gas surface density in a set of nearby galaxies. These authors proposed that the main driver of this correlation is the turbulent pressure of gas, while the dynamical state of the gas (e.g.\ degree of self-gravity) would explain the scatter in the direction perpendicular to the correlation. Thus, if linewidths are mainly reflecting a dependence on molecular gas pressure, and if this internal pressure is associated with the dynamical equilibrium pressure expected across the disc (e.g.\ \citealt{2013ApJ...779...46H}; \citealt{2016IAUS..315...30H}; \citealt{schruba19}; Sun et al.~in prep),
linewidths should correlate with galactocentric radius. In that case, it would not be surprising to obtain a correlation that goes in the same sense as that found with stellar mass surface density (i.e.\ an anti-correlation between SFE and $\sigma$) as both should, to first approximation, decrease with galactocentric radius.

The fact that we find velocity dispersion values for HCN which are comparable to the ones for CO confirms that we are not resolving the inner turbulent motions within individual clumps (which is expected given our ${\sim} 100$\,pc resolution). Instead, HCN linewidths must be primarily capturing the velocity dispersion among dense clumps, reflecting a mix of dynamical streaming motions and larger-scale turbulence in the gaseous disc. This connects with the role of 
dynamical environment, which we will discuss in the next sub-section.

\subsection{Suppressed star formation in M51 despite high dense gas fractions} 
\label{DiscussSouth}

Our measurements indicate that the southern arm seems particularly inefficient at producing stars when compared to regions that lie at similar stellar mass surface densities. This had already been pointed out by \citet{2013ApJ...779...45M} and \citet{2017ApJ...846...71L} referenced to the molecular gas traced by CO (i.e.\ $\mathrm{SFE_{mol}}$). Our results show that the long depletion times associated with the southern arm region are not due to a lack of dense gas, but instead result from a lower star formation efficiency of the dense gas. While the southern arm is a strong lower outlier in the $\mathrm{SFE_{dense}}-\Sigma_{\star}$ plot, it follows the general trend when 
examined against the velocity dispersion of HCN ($\mathrm{SFE_{dense}-\sigma_{HCN}}$). This might indicate that the dynamical environment is playing a role in stabilising these clouds against collapse (e.g.\ as a result of shear, see also \citealt{2018A&A...609A..60K}), as pointed out by \citet{2013ApJ...779...45M} and \citet{2018ApJ...854..100M}. Since $\mathrm{SFE_{dense}}$ is suppressed (and not only $\mathrm{SFE_{mol}}$), we might be witnessing the effects of dynamical stabilisation of dense gas at play in the southern spiral arm of M51. This is a result that uniquely comes from having fine enough resolution in HCN to distinguish distinct environments, like we do here.

\referee{Overall, our results suggest that} a high dense gas fraction is not enough to sustain star formation. The regions that are converting the available molecular gas into stars most efficiently are not those that have the highest dense gas fractions, but quite the opposite. We note that a similar trend was found by \citet{2015ApJ...813..118M} analysing dense gas and star formation in NGC\,3627 at similar scales as this work (${\sim}300$\,pc); they also found lower $\mathrm{SFE_{dense}}$ associated with higher $F_\mathrm{dense}$.

\subsection{Dense gas fraction and dense gas surface density as predictors of star formation} 

We have also considered the recent result from \citet{2018MNRAS.475.5550V}, who found a linear correlation between SFRs and dense gas fraction at ${\sim} 100$\,pc resolution in M31. When we combine the measurements from \citet{2017ApJ...836..101C} for the outer arm in M51 with our new data, we recover a significant correlation between the dense gas fraction and SFR surface density ($\rho_{\mathrm{\Sigma_{SFR}-F_{dense}}}=0.63$). If we simultaneously consider the literature measurements for M31 and NGC\,3627, the resulting correlation coefficient becomes $\rho_{\mathrm{\Sigma_{SFR}-F_{dense}}}=0.66$. Most likely, this correlation is indirectly capturing the fact that higher gas surface densities tend to result in higher star formation rate surface densities. For example, $F_\mathrm{dense}$ is known to strongly correlate with $\Sigma_\mathrm{mol}$ \citep[e.g.][]{2018ApJ...868L..38G}, and higher $\Sigma_\mathrm{mol}$ naturally leads to higher $\Sigma_\mathrm{SFR}$ on average  \citep[e.g.][]{2013AJ....146...19L}.

This result can also be connected with the anti-correlation that we found between $\mathrm{SFE_{dense}}$ and $\Sigma_\star$.
$\mathrm{SFE_{dense}}$ is, by construction, proportional to $\Sigma_\mathrm{SFR} / \mathrm{HCN}$. Thus, the correlation between $\Sigma_\mathrm{SFR}$ and HCN/CO suggests that $\mathrm{SFE_{dense}}$ also correlates with 1/CO (after dividing both terms of the correlation by HCN). This second correlation is expected, because CO intensity roughly increases towards the centre of galaxies, and so does the stellar mass surface density. Therefore, the correlation between $\Sigma_\mathrm{SFR}$ and $F_\mathrm{dense}$ can be regarded to some extent as indirectly equivalent to the anti-correlation that we found between $\mathrm{SFE_{dense}}$ and $\Sigma_\star$.

On the other hand, when we consider all the ${\sim} 100$\,pc-resolution data available to us, the correlation that we find between $\Sigma_\mathrm{SFR}$ and the dense gas surface density, $\mathrm{\Sigma_{dense}}$, is even stronger than the correlation found with the dense gas fraction ($\rho_{\mathrm{\Sigma_{SFR}-\Sigma_{dense}}}=0.74$ instead of $\rho_{\mathrm{\Sigma_{SFR}}-F_\mathrm{dense}}=0.66$). \citet{2018MNRAS.475.5550V} also found a comparable degree of correlation between $\Sigma_\mathrm{SFR}$ and $\mathrm{\Sigma_{dense}}$ ($\rho = 0.62$) as between $\Sigma_\mathrm{SFR}$ and $F_\mathrm{dense}$ in their data for M31, but discussed that the correlation with dense gas fraction increased dramatically (to $\rho = 0.98$) when excluding a specific outlier. Our data do not support that exceptionally strong correlation, but confirm indeed that the trend with $F_\mathrm{dense}$ is significant and comparable to the one with the dense gas surface density.




\subsection{Does the star formation law ``break down'' at 100\,pc scales?} 
\label{Sec:SFbreakdown}

 In Sect.~\ref{Sec:offsets} we have shown that the peaks in 33\,GHz, HCN, and CO emission are not always perfectly aligned. We remind the reader that young stars are traced by 33\,GHz free-free emission from the ionised gas  surrounding them (\ion{H}{II} regions). Sometimes, we find HCN and CO maxima which do not correspond to a peak in star formation; this is not too surprising, as a cloud bright in CO and HCN could be dynamically stabilised and not have (yet) begun to collapse to form stars. We also encounter the opposite case, strong 33\,GHz peaks that do not have any significant CO or HCN clumps associated with them; this could be reflecting the fact that a vigorous episode of star formation (probably resulting in young star clusters) has triggered feedback effects which have destroyed the entire parent cloud. However, in most of the cases, we find intermediate situations where the gas and star formation tracers seem to be coupled, but show some small offsets (typically up to ${\sim} 2'' \approx 80$\,pc). For CO and 33\,GHz, where we benefit from a higher resolution (${\sim} 1''$) and larger field of view ($11 \times 7$\,kpc), we have quantified the offsets between the strongest 33\,GHz peaks and the GMCs identified by PAWS \citep{2014ApJ...784....3C}. We find that 85\% of the 33\,GHz peaks have a GMC at a (projected) distance smaller than $2''$, with a median (projected) offset of $1.2''$.

What we see agrees with \citet{2009ApJS..184....1K},  who identified different situations in the LMC regarding the connection between molecular clouds and young stellar clusters. First of all, it is possible to find gas which has not (yet) collapsed to form massive stars and which corresponds to isolated GMCs. We certainly find cases like this in our observations, as we illustrated in Fig.~\ref{fig:offsets}. For some time, while star formation proceeds, newly formed stars and their parent clouds can coexist; they would appear as aligned or nearby peaks in our observations (e.g.\ because part of a cloud collapses, while a neighbouring part survives), a situation that we also encounter. Finally, depending on what fraction of a given cloud is converted into stars, and the timescales and efficiency of stellar feedback on its parent cloud, the molecular gas can fully disappear (so we only see the product of star formation; e.g.\ \citealt{2017MNRAS.470.4453R}), or result in spatial offsets between the clouds and the \ion{H}{II} regions or clusters. In addition, offsets of up to ${\sim} 100$\,pc between emission peaks of HCN and SFR tracers have also been reported for the barred galaxies NGC\,7522 \citep{2013ApJ...768...57P} and NGC\,3627 \citep{2015ApJ...813..118M}. 


A practical consequence of this spatial decoupling is that we obtain different depletion times depending on where we place the apertures (both for the dense and bulk molecular gas), as shown by Table~\ref{table:tdep}. In agreement with \citet{2010ApJ...722.1699S} and \citet{kruijssen19}, we find the most extreme differences between placing the apertures on the peaks of molecular gas (either CO or HCN) and the star formation maxima (33\,GHz).
Unlike, for example, FUV or TIR, with 33\,GHz we are tracing very recent star formation (timescales of ${\lesssim} 10$\,Myr). This means that our chances of capturing young \ion{H}{II} regions (mapped by 33\,GHz) next to their gaseous cocoon are not negligible; as long as the separation remains below our beam size, the two phases will be captured together in the same aperture. In this case, the differences that we measure in depletion times are expected to be driven mostly by the evolution of individual star-forming regions \citep{2014MNRAS.439.3239K,2018MNRAS.479.1866K}.
For this reason, individual aperture measurements are not expected to be representative of the time-averaged star formation process; on the contrary, they are stochastic in the sense that they reflect a snapshot of the star formation cycle (or the average of a small number of star-forming regions which can be at different stages of the process). Only as an  ensemble (for example, each of the different areas that we target in M51) can the measurements be representative of the time-averaged star formation process as observed at lower resolutions.




\section{Summary and conclusions} 
\label{Sec:concl}

We have presented 33\,GHz continuum observations of the disc of M51 with the VLA, tracing star formation at ${\sim} 100$\,pc scales through free-free emission. We have combined this map with three specific pointings of \mbox{HCN(1-0)} obtained with NOEMA, and the public \mbox{CO(1-0)} map from PAWS, all at a matched resolution of $3'' \sim 100$\,pc. The aperture measurements can be found in Table~\ref{table:results}. Our main results are the following:

\begin{enumerate}

\item Our measurements fill the gap between Galactic cores and larger extragalactic structures on the HCN-TIR plane. However, the scatter shows systematic variations in the form of vertical offsets between the regions that we targeted (Fig.~\ref{fig:TIR-HCN}).

\item Our high-resolution maps reveal spatial offsets between peaks in 33\,GHz (recent star formation sites), HCN (dense gas), and CO (bulk molecular gas). This is expected as a result of the time evolution of star-forming regions (Fig.~\ref{fig:offsets}).

\item Far from being approximately constant, as required by density threshold models, the star formation efficiency of the dense gas ($\mathrm{SFE_{dense}}$) varies by more than 1\,dex among the three regions that we targeted (Fig.~\ref{fig:ICA-SFEdense}). Moreover, $\mathrm{SFE_{dense}}$ and dense gas fraction ($F_\mathrm{dense}$) correlate with stellar mass surface density ($\Sigma_\star$), in agreement with previous studies at lower resolution (Fig.~\ref{fig:ICA-SFEdense}). \referee{The correlations are stronger when we expand the dynamic range by including other 100\,pc-scale measurements from the literature.}

\item Star formation is suppressed in the southern spiral arm in spite of high dense gas fractions (with a clearly lower $\mathrm{SFE_{dense}}$). This, together with the previous point, suggests that the local dynamical environment plays a special role modulating star formation.

\item We find an anti-correlation between $\mathrm{SFE_{dense}}$ and velocity dispersion ($\sigma_\mathrm{HCN}$). Specifically, the southern arm has significantly higher $\sigma_\mathrm{HCN}$ than the northern spurs, while $\Sigma_\star$ is similar; the increase in $\sigma_\mathrm{HCN}$ is likely associated with the suppression of star formation (Fig.~\ref{fig:Sigma-SFE}).

\item We confirm a significant correlation between SFR surface density and $F_\mathrm{dense}$, as pointed out by \citet{2018MNRAS.475.5550V}, when we simultaneously consider our results and literature measurements (Fig.~\ref{fig:Fdense-SFR}). However, the correlation is not stronger than the one found between SFR and dense gas surface density. The correlation can be expected given the trend that we found between $\mathrm{SFE_{dense}}$ and $\Sigma_\star$, and as an indirect consequence of more molecular gas resulting in higher star formation.

\end{enumerate}

In conclusion, our observations suggest that the presence of dense gas is not sufficient to sustain high levels of star formation. Specifically, galactic environment seems to play a prominent role in modulating how efficiently dense gas transforms into stars.

\small  
%
\begin{acknowledgements}   
Based on observations carried out with the IRAM Interferometer NOEMA. IRAM is supported by INSU/CNRS (France), MPG (Germany), and IGN (Spain).
The National Radio Astronomy Observatory is a facility of the National Science Foundation operated under cooperative agreement by Associated Universities, Inc.
This work was carried out as part of the PHANGS collaboration. We would like to thank Kazimierz Sliwa, Ga\"{e}lle Dumas, Melanie Krips, and J\"{u}rgen Ott for helpful assistance regarding data reduction. We also thank Neven Tomi\v{c}i\'{c} for providing SFR measurements for the Andromeda galaxy, and Sean T.\ Linden, Eve C.\ Ostriker, and S\'{e}bastien Viaene for useful feedback. \referee{The authors would also like to thank the anonymous referee for constructive comments.}
MQ and SEM acknowledge (partial) funding from the Deutsche Forschungsgemeinschaft (DFG) via grant SCHI\,536/7-2 as part of the priority programme SPP 1573 “ISM-SPP: Physics of the Interstellar Medium”.
ES and CMF acknowledge funding from the European Research Council (ERC) under the European Union’s Horizon 2020 research and innovation programme (grant agreement No. 694343).
The work of AKL and DU is partially supported by the National Science Foundation under Grants No. 1615105, 1615109, and 1653300.
FB acknowledges funding from the European Research Council (ERC) under the European Union’s Horizon 2020 research and innovation programme (grant agreement No. 726384).
APSH is a fellow of the International Max Planck Research School for Astronomy and Cosmic Physics at the University of Heidelberg (IMPRS-HD).
The work of MJG and DU is partially supported by the National Science Foundation under Grants No. 1615105, 1615109, and 1653300.
JMDK and MC gratefully acknowledge funding from the Deutsche Forschungsgemeinschaft (DFG) in the form of an Emmy Noether Research Group (grant number KR4801/1-1). JMDK gratefully acknowledges funding from the European Research Council (ERC) under the European Union’s Horizon 2020 research and innovation programme via the ERC Starting Grant MUSTANG (grant agreement number 714907).
SGB acknowledges support from the Spanish MINECO grant
AYA2016-76682-C3-2-P.
JP acknowledges support from the Programme National “Physique et Chimie du Milieu Interstellaire” (PCMI) of CNRS/INSU with INC/INP co-funded by CEA and CNES.
ER acknowledges the support of the Natural Sciences and Engineering Research Council of Canada (NSERC), funding reference number RGPIN-2017-03987.
SCOG acknowledges support from the DFG via SFB 881 ``The Milky Way System'' (sub-projects B1, B2 and B8).
\end{acknowledgements}

\normalsize

\appendix
\section{Comparison with other SFR tracers}
\label{sec:SFRcomparison}

Here we compare the SFRs derived from 33\,GHz with other tracers of common use in the literature. Our goal is not to provide a systematic bench-marking across calibrations based on different tracers and spatial scales, but simply to derive a zeroth-order rescaling in order to homogenise the measurements from the literature with our new datapoints, so that we avoid systematic offsets when plotting them together. While 33\,GHz is, in principle, a more direct tracer of the ionising radiation associated with the star formation process (bypassing the uncertainties associated with extinction), there are some concerns associated with its use; we discuss such caveats in greater detail in Sect.~\ref{Sec:usefultracer}.

\begin{figure*}[t]
\begin{center}
\includegraphics[trim=0 295 100 0, clip,width=1.0\textwidth]{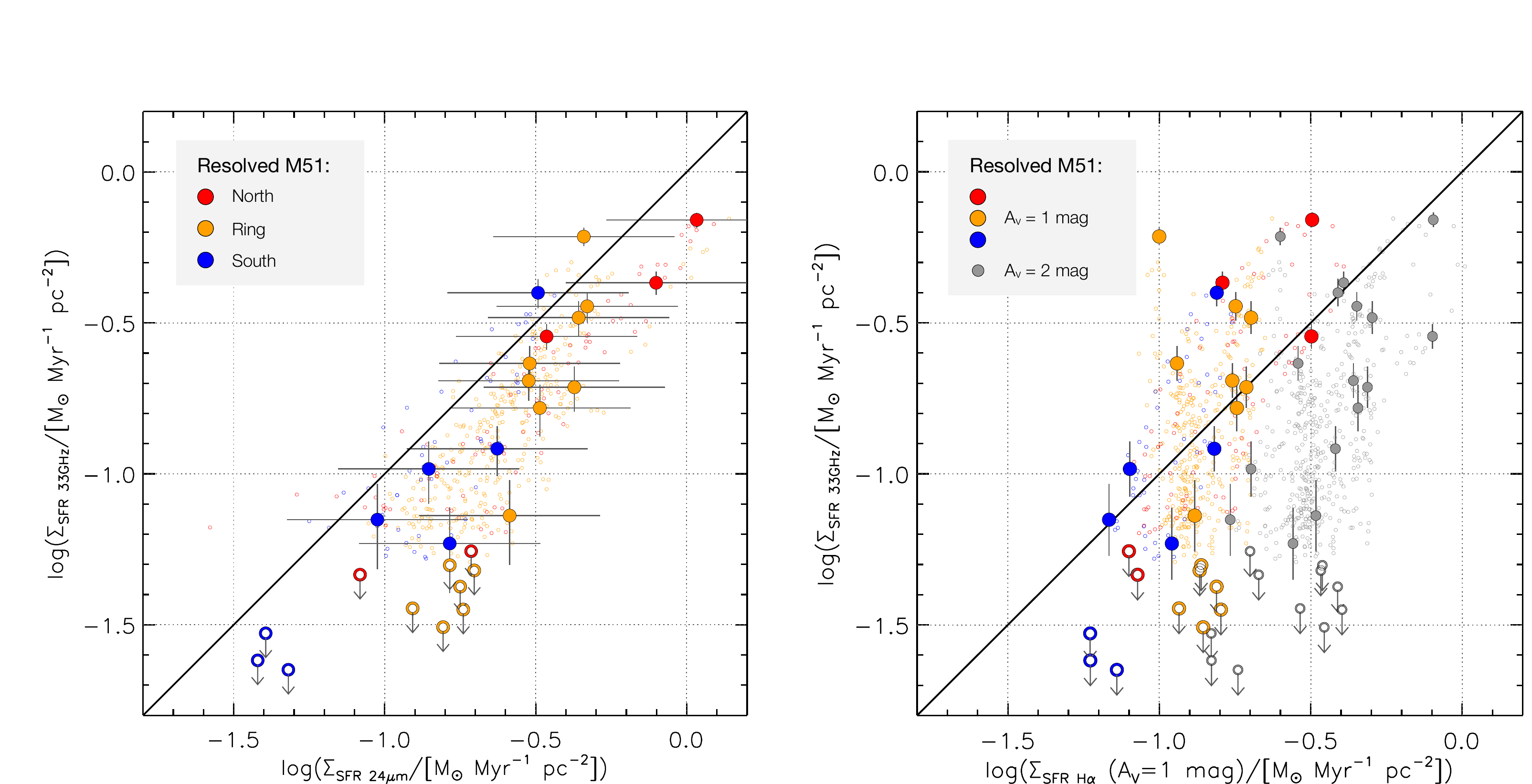}
\end{center}
\caption{\textit{Left panel:} SFR based on 33\,GHz vs SFR based on 24\,$\mu$m (following the recipe from \citealt{2007ApJ...666..870C}), for $3''$ apertures (${\sim} 100$\,pc).
\textit{Right panel:} SFR based on 33\,GHz vs SFR from H$\alpha$ (coloured circles assuming an extinction of 1\,mag; gray points show the effect of 2\,mag extinction). Since the uncertainty is dominated by the effect of extinction (illustrated by the offset between coloured and gray points), we do not explicitly plot error bars for the H$\alpha$-based SFRs. In both cases, the filled circles correspond to apertures centred on HCN peaks, while 
the small open circles in the background show all detections (above $3\sigma$) for Nyquist-sampled apertures across our fields of view. Downward arrows indicate upper limits (in the SFRs based on 33\,GHz).
}
\label{fig:appendixSFRtirHalpha-SFR33}
\end{figure*}

The left panel of Fig.~\ref{fig:appendixSFRtirHalpha-SFR33} shows how our 33\,GHz-based SFRs compare with the SFRs derived from 24\,$\mu$m. We can perform this comparison at a matched resolution of $3'' {\sim} 100$\,pc (which defines the main apertures throughout the paper) using the deconvolved PSF of the \textit{Spitzer} 24\,$\mu$m image from \citet{2011AJ....141...41D}. If we follow the empirical recipe from \citet{2007ApJ...666..870C}, shown as the coloured circles in the plot, the average ratio for the same set of apertures centred on HCN peaks is SFR(33\,GHz) / SFR(24\,$\mu$m-Calzetti) $= 0.70 \pm 0.28$. We note that \citet{2007ApJ...666..870C} adopted the same IMF as \citet{2011ApJ...737...67M}, so the offset is not due to a different choice of IMF. The small open circles in the background show the relationship between the two tracers for all of our significant detections (above $3\sigma$) for Nyquist-sampled apertures across our fields of view, which highlights the considerable point-to-point scatter. As a side note, if instead of using the empirical prescription from \citet{2007ApJ...666..870C} we first convert to TIR  following \citet{2013MNRAS.431.1956G} and then translate these luminosities into SFRs according to \citet{2011ApJ...737...67M}, we would obtain an even larger discrepancy, with an average SFR(33\,GHz) / SFR(24\,$\mu$m-TIR) ratio of $0.37 \pm 0.34$ for the apertures centred on HCN peaks. However, local variations in the 24\,$\mu$m-to-TIR ratio are expected within M51 (see Fig.~12 in \citealt{2018ApJ...858...90G}).

The right panel of Fig.~\ref{fig:appendixSFRtirHalpha-SFR33} confirms that the 33\,GHz-based SFRs are in agreement with the SFRs implied by H$\alpha$ for a typical extinction of $1-2$\,mag; this is a reasonable value for these areas of M51 according to the extinction map from the VENGA survey, which partially covered our target \citep{2009ApJ...704..842B}. In any case, the scatter is clearly very large.

\citet{2017ApJ...836..101C} used TIR as their SFR tracer, extrapolated from \textit{Herschel}/PACS 70\,$\mu$m fluxes. To convert their TIR to SFR we follow \citet{2011ApJ...737...67M}. Given the systematic differences among SFR tracers that we just discussed, we rescaled the SFRs from \citet{2017ApJ...836..101C} by an empirically derived factor (0.42) to enforce consistency. 
Since \citet{2017ApJ...836..101C} focussed on HCN-rich regions (their apertures are limited to areas of sufficiently bright HCN emission), we derive this empirical rescaling factor in M51 using our nominal apertures centred on HCN peaks.
For these apertures, we calculated TIR-based SFRs using 70\,$\mu$m and compared against SFRs based on 33\,GHz, after convolving our 33\,GHz map to the PSF from \textit{Herschel}/PACS \citep[FWHM$=5.7''$;][]{2013ApJ...779...42S}; we performed the convolution using the kernels from \citet{2011PASP..123.1218A}. The average ratio found this way is SFR(33\,GHz) / SFR(70\,$\mu$m-TIR) $= 0.42 \pm 0.13$.

We emphasise that for M31 we do not plot the SFRs from \citet{2013ApJ...769...55F}, which were used by \citet{2018MNRAS.475.5550V}, but the values derived by \citet{2019ApJ...873....3T}, who calibrated an improved FUV$+$24\,$\mu$m hybrid prescription  using extinction-corrected H$\alpha$:

\begin{equation}
\Sigma_\mathrm{SFR}\,\mathrm{(FUV + IR)} = 4.42 \times 10^{-44} \times [\Sigma\mathrm{(FUV)} + (27 \pm 4) \times \Sigma\mathrm{(IR)}],
\end{equation}

\noindent
which results in $M_\odot \, \mathrm{yr^{-1}}\, \mathrm{kpc^{-1}}$ if $\Sigma\mathrm{(FUV)}$ and $\Sigma\mathrm{(IR)}$ are expressed in $\mathrm{erg}\, \mathrm{s}^{-1}\, \mathrm{kpc^{-1}}$.
According to this new hybrid recipe, for the apertures from \citet{2005A&A...429..153B} and \citet{2018MNRAS.475.5550V} in M31, the SFRs are strongly dominated by the 24\,$\mu$m component, with FUV typically contributing only ${\sim} 1\%$. This is why, to bring the measurements in M31 to the same scale as our 33\,GHz SFRs in M51, we only consider the rescaling factor associated with 24\,$\mu$m (the factor 0.70 discussed above, which is the average ratio of the SFRs following \citealt{2007ApJ...666..870C} and based on 33\,GHz). On the other hand, for NGC\,3627, since \citet{2015ApJ...813..118M} used the same tracer that we employ here (33\,GHz radio continuum), we do not apply any offsets to those datapoints.

\begin{figure}[t]
\begin{center}
\includegraphics[trim=0 360 590 0, clip,width=0.48\textwidth]{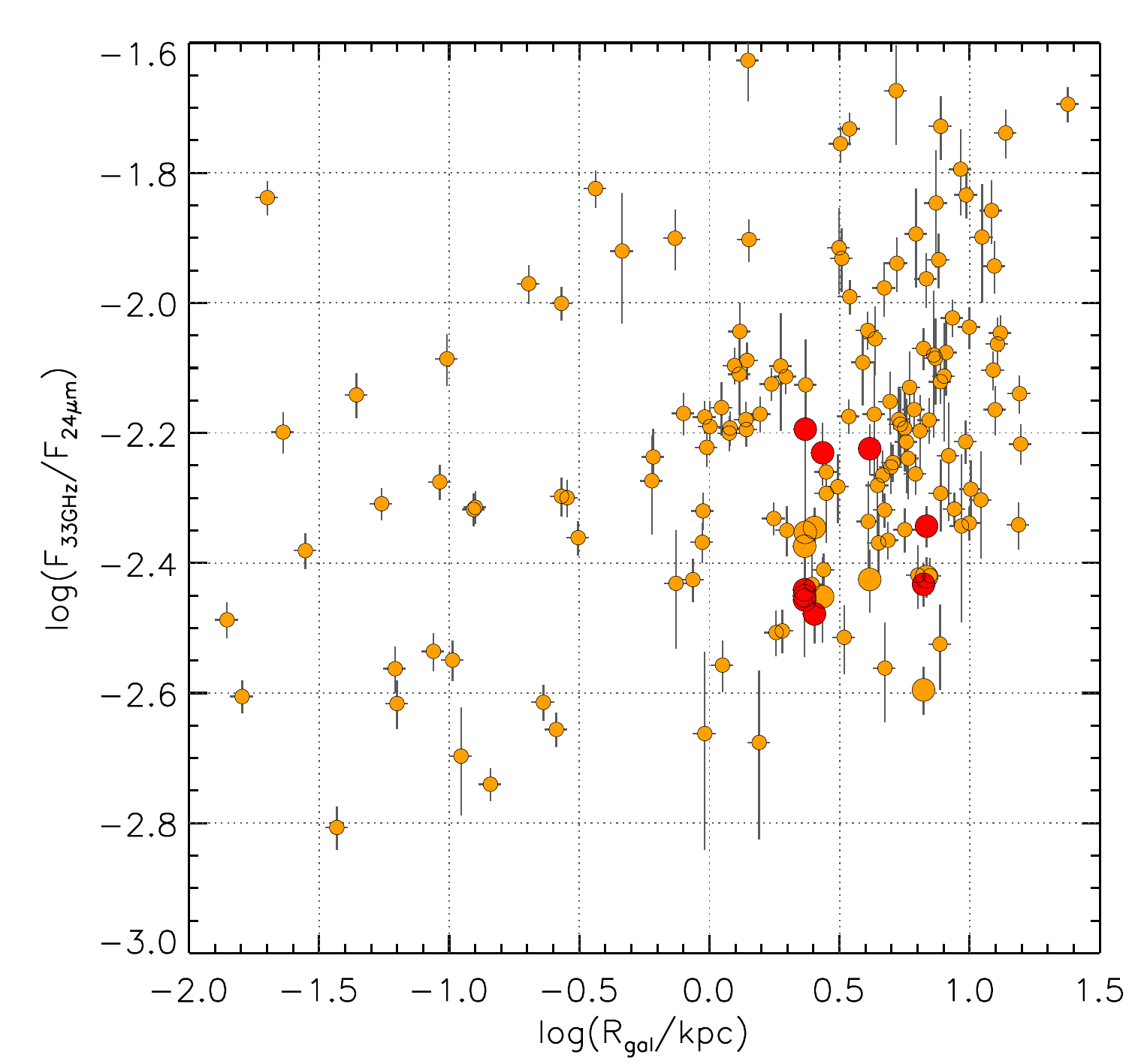}
\end{center}
\caption{33\,GHz-to-24\,$\mu$m flux ratio as a function of galactocentric radius at $7''$ resolution from \citet{2018ApJS..234...24M}, shown as small diamonds. These apertures include regions of M51 that overlap with our measurements, and we highlight those ratios as large circles (orange for the Murphy et al.~values and red for the analogous measurements using our maps).}
\label{fig:33-to-24_Murphy}
\end{figure}

To set in context the differences between SFRs based on 33\,GHz and those obtained from 24\,$\mu$m, in Fig.~\ref{fig:33-to-24_Murphy} we plot the 33\,GHz-to-24\,$\mu$m flux ratio as a function of galactocentric radius. We compare individual apertures across nearby galaxies from \citet{2018ApJS..234...24M} and, for the positions in M51 which overlap with their study, we highlight the ratios obtained using our own maps (after convolving to the same spatial resolution of $7''$). We find that our measurements in M51 are not exactly coincident with the same apertures by \citet{2018ApJS..234...24M}, but they are compatible within the large error bars. Moreover, as an ensemble, the ratios in M51 lie on the area of the parameter space covered by other nearby galaxies, but with a tendency for lower 33\,GHz-to-24\,$\mu$m ratios.
Even though they are globally consistent, we do not expect a perfect one-to-one match between each of our measurements and those from \citet{2018ApJS..234...24M} due to noise fluctuations, the slightly different $uv$ coverage of the observations, or different cleaning strategies.

\bibliography{/Users/mquereje/Documents/E1_SCIENCE_UTIL/mq}{}
\bibliographystyle{aa}{}

\clearpage
\onecolumn
 \begin{landscape}
 
 \normalsize

\begin{table*}[t!]
\begin{center}
\caption[h!]{Results for 3$''$ apertures in M51 centred on HCN peaks.}
\setlength{\tabcolsep}{5pt} 
\begin{tabular}{lcc
r@{\hspace{0.5\tabcolsep}}C{0.2cm}@{\hspace{0.5\tabcolsep}}l
r@{\hspace{0.5\tabcolsep}}C{0.2cm}@{\hspace{0.5\tabcolsep}}l
r@{\hspace{0.5\tabcolsep}}C{0.2cm}@{\hspace{0.5\tabcolsep}}l
r@{\hspace{0.5\tabcolsep}}C{0.2cm}@{\hspace{0.5\tabcolsep}}l
R{0.8cm}@{\hspace{0.5\tabcolsep}}C{0.2cm}@{\hspace{0.5\tabcolsep}}L{0.8cm}
r@{\hspace{0.5\tabcolsep}}C{0.2cm}@{\hspace{0.5\tabcolsep}}l
r@{\hspace{0.5\tabcolsep}}C{0.2cm}@{\hspace{0.5\tabcolsep}}l
r@{\hspace{0.5\tabcolsep}}C{0.2cm}@{\hspace{0.5\tabcolsep}}l
r@{\hspace{0.5\tabcolsep}}C{0.2cm}@{\hspace{0.5\tabcolsep}}l}
\hline\hline
Region & RA$\rm_{J2000}$ & DEC$\rm_{J2000}$ 
& \multicolumn{3}{c}{$\Sigma_\star$} 
& \multicolumn{3}{c}{$L_\mathrm{TIR}$} 
& \multicolumn{3}{c}{$S_\nu^\mathrm{33\,GHz}$}
& \multicolumn{3}{c}{$f_{\rm T}^\mathrm{33\,GHz}$} 
& \multicolumn{3}{c}{$\Sigma_\mathrm{SFR}$} 
& \multicolumn{3}{c}{$I_\mathrm{HCN}$} 
& \multicolumn{3}{c}{$I_\mathrm{CO}$} 
& \multicolumn{3}{c}{$\sigma_\mathrm{HCN}$} 
& \multicolumn{3}{c}{$\sigma_\mathrm{CO}$} \\
 & (deg) & (deg) 
 & \multicolumn{3}{c}{($\mathrm{M_\odot / pc^2})$} 
 & \multicolumn{3}{c}{($10^6 L_\odot$)} 
 & \multicolumn{3}{c}{($\mu$Jy)}  
 & \multicolumn{3}{c}{(\%)}  
 & \multicolumn{3}{c}{(M$_\odot$/Myr/pc$^{2}$)} 
 & \multicolumn{3}{c}{(K\,km\,s$^{-1}$)} 
 & \multicolumn{3}{c}{(K\,km\,s$^{-1}$)} 
 & \multicolumn{3}{c}{(km\,s$^{-1}$)} 
 & \multicolumn{3}{c}{(km\,s$^{-1}$)} \\
 \hline
  M51 north     &  202.4564     &  47.21080     &        408&$\pm$&       131     &      99.7&$\pm$&     25.8     &     102.0&$\pm$&      5.4     &      78.5&$\pm$&      4.8     &      0.43&$\pm$&     0.04     &      4.88&$\pm$&     0.70     &     122.3&$\pm$&     12.2     &      12.1&$\pm$&      2.0     &     12.73&$\pm$&     0.40 \\
  M51 north     &  202.4691     &  47.21020     &        432&$\pm$&       134     &       8.7&$\pm$&      2.3     &      \multicolumn{3}{c}{$<$13.3} &      23.1&$\pm$&     25.3     &      \multicolumn{3}{c}{$<$0.05} &      2.92&$\pm$&     0.59     &      66.1&$\pm$&      6.6     &      11.0&$\pm$&      2.7     &     11.08&$\pm$&     0.66 \\
  M51 north     &  202.4627     &  47.21120     &        416&$\pm$&       130     &      21.8&$\pm$&      5.6     &      17.1&$\pm$&      4.5     &      47.1&$\pm$&     15.7     &      \multicolumn{3}{c}{$<$0.06} &      2.65&$\pm$&     0.57     &      53.5&$\pm$&      5.3     &      10.1&$\pm$&      2.5     &      7.83&$\pm$&     0.43 \\
  M51 north     &  202.4669     &  47.21200     &        383&$\pm$&       124     &     139.1&$\pm$&     36.0     &     140.1&$\pm$&      6.2     &      92.3&$\pm$&      1.9     &      0.69&$\pm$&     0.04     &      3.68&$\pm$&     0.66     &      83.6&$\pm$&      8.4     &       7.8&$\pm$&      1.5     &      9.22&$\pm$&     0.39 \\
  M51 north     &  202.4676     &  47.21300     &        378&$\pm$&       122     &      40.5&$\pm$&     10.5     &      62.4&$\pm$&      5.0     &      85.1&$\pm$&      3.9     &      0.29&$\pm$&     0.03     &      3.69&$\pm$&     0.64     &      46.2&$\pm$&      4.6     &      10.8&$\pm$&      2.2     &      6.83&$\pm$&     0.52 \\
   M51 ring     &  202.4642     &  47.20000     &       1686&$\pm$&       509     &      56.6&$\pm$&     14.6     &      95.8&$\pm$&      5.1     &      69.8&$\pm$&      6.6     &      0.36&$\pm$&     0.04     &     12.15&$\pm$&     1.32     &     209.5&$\pm$&     20.9     &      16.7&$\pm$&      1.0     &     14.29&$\pm$&     0.29 \\
   M51 ring     &  202.4677     &  47.20020     &       1744&$\pm$&       526     &      38.3&$\pm$&      9.9     &      54.0&$\pm$&      4.6     &      57.1&$\pm$&      9.7     &      0.17&$\pm$&     0.03     &      7.02&$\pm$&     0.85     &     104.3&$\pm$&     10.4     &      13.1&$\pm$&      1.2     &      8.28&$\pm$&     0.28 \\
   M51 ring     &  202.4682     &  47.20070     &       1324&$\pm$&       401     &      50.9&$\pm$&     13.2     &      59.0&$\pm$&      4.6     &      61.2&$\pm$&      8.8     &      0.19&$\pm$&     0.03     &      7.35&$\pm$&     0.91     &     118.7&$\pm$&     11.9     &      14.5&$\pm$&      1.4     &      9.86&$\pm$&     0.33 \\
   M51 ring     &  202.4713     &  47.19100     &       2649&$\pm$&       797     &      13.5&$\pm$&      3.5     &      \multicolumn{3}{c}{$<$12.7} &      35.0&$\pm$&     17.6     &      \multicolumn{3}{c}{$<$0.04} &      4.85&$\pm$&     0.63     &      72.8&$\pm$&      7.3     &      16.3&$\pm$&      1.9     &      8.06&$\pm$&     0.40 \\
   M51 ring     &  202.4718     &  47.19150     &       3129&$\pm$&       941     &      17.3&$\pm$&      4.5     &      \multicolumn{3}{c}{$<$12.7} &      42.1&$\pm$&     17.2     &      \multicolumn{3}{c}{$<$0.03} &      5.40&$\pm$&     0.69     &      58.2&$\pm$&      5.8     &      15.1&$\pm$&      1.6     &      7.82&$\pm$&     0.45 \\
   M51 ring     &  202.4740     &  47.19230     &       3221&$\pm$&       968     &      20.4&$\pm$&      5.3     &      \multicolumn{3}{c}{$<$12.7} &      47.9&$\pm$&     17.1     &      \multicolumn{3}{c}{$<$0.04} &      5.27&$\pm$&     0.68     &      71.4&$\pm$&      7.1     &      16.1&$\pm$&      1.8     &     10.87&$\pm$&     0.56 \\
   M51 ring     &  202.4736     &  47.19280     &       3825&$\pm$&      1149     &      19.9&$\pm$&      5.2     &      \multicolumn{3}{c}{$<$12.7} &      58.7&$\pm$&     14.2     &      \multicolumn{3}{c}{$<$0.04} &      5.28&$\pm$&     0.68     &      77.9&$\pm$&      7.8     &      17.5&$\pm$&      2.1     &     12.12&$\pm$&     0.69 \\
   M51 ring     &  202.4591     &  47.19480     &        607&$\pm$&       188     &      35.3&$\pm$&      9.1     &      65.4&$\pm$&      4.6     &      66.2&$\pm$&      7.7     &      0.23&$\pm$&     0.03     &      4.79&$\pm$&     0.75     &     183.1&$\pm$&     18.3     &      11.9&$\pm$&      2.0     &     18.25&$\pm$&     0.53 \\
   M51 ring     &  202.4728     &  47.19020     &       1712&$\pm$&       516     &      29.9&$\pm$&      7.7     &      28.4&$\pm$&      4.3     &      47.8&$\pm$&     12.8     &      0.07&$\pm$&     0.02     &      6.22&$\pm$&     0.81     &     115.4&$\pm$&     11.5     &      14.2&$\pm$&      1.5     &     10.65&$\pm$&     0.32 \\
   M51 ring     &  202.4726     &  47.19080     &       2484&$\pm$&       747     &      22.4&$\pm$&      5.8     &      15.2&$\pm$&      4.2     &      45.3&$\pm$&     14.8     &      \multicolumn{3}{c}{$<$0.05} &      5.17&$\pm$&     0.67     &      90.4&$\pm$&      9.0     &      13.5&$\pm$&      1.4     &      9.72&$\pm$&     0.35 \\
   M51 ring     &  202.4760     &  47.19170     &       1816&$\pm$&       547     &      18.3&$\pm$&      4.7     &      17.0&$\pm$&      4.2     &      54.4&$\pm$&     11.8     &      \multicolumn{3}{c}{$<$0.05} &      4.60&$\pm$&     0.62     &      64.0&$\pm$&      6.4     &      20.1&$\pm$&      2.9     &      9.65&$\pm$&     0.46 \\
   M51 ring     &  202.4581     &  47.19180     &        815&$\pm$&       249     &      55.0&$\pm$&     14.2     &     136.7&$\pm$&      5.8     &      83.3&$\pm$&      3.7     &      0.61&$\pm$&     0.04     &      7.41&$\pm$&     1.09     &     177.4&$\pm$&     17.7     &      16.6&$\pm$&      2.6     &     14.34&$\pm$&     0.35 \\
   M51 ring     &  202.4620     &  47.19870     &       1183&$\pm$&       360     &      52.7&$\pm$&     13.7     &      93.5&$\pm$&      5.1     &      65.6&$\pm$&      7.5     &      0.33&$\pm$&     0.04     &      4.81&$\pm$&     0.68     &     122.2&$\pm$&     12.2     &      13.9&$\pm$&      1.8     &     15.35&$\pm$&     0.55 \\
   M51 ring     &  202.4659     &  47.19950     &       1624&$\pm$&       491     &      35.0&$\pm$&      9.1     &      55.8&$\pm$&      4.6     &      68.0&$\pm$&      7.4     &      0.20&$\pm$&     0.03     &      6.28&$\pm$&     0.79     &     110.3&$\pm$&     11.0     &      11.5&$\pm$&      1.1     &     10.86&$\pm$&     0.40 \\
  M51 south     &  202.4613     &  47.18620     &        548&$\pm$&       168     &      15.4&$\pm$&      4.0     &      34.4&$\pm$&      4.2     &      56.2&$\pm$&     10.9     &      0.10&$\pm$&     0.02     &      8.91&$\pm$&     1.01     &     147.8&$\pm$&     14.8     &      18.7&$\pm$&      1.7     &     14.93&$\pm$&     0.42 \\
  M51 south     &  202.4586     &  47.19040     &        790&$\pm$&       242     &      37.8&$\pm$&      9.8     &     105.2&$\pm$&      5.2     &      70.5&$\pm$&      6.5     &      0.40&$\pm$&     0.04     &     16.80&$\pm$&     1.87     &     221.6&$\pm$&     22.2     &      20.0&$\pm$&      1.7     &     15.09&$\pm$&     0.29 \\
  M51 south     &  202.4667     &  47.18170     &        478&$\pm$&       147     &      18.3&$\pm$&      4.7     &      20.3&$\pm$&      4.1     &      54.0&$\pm$&     12.7     &      0.06&$\pm$&     0.02     &      6.59&$\pm$&     0.81     &     177.9&$\pm$&     17.8     &      18.2&$\pm$&      2.2     &     16.20&$\pm$&     0.36 \\
  M51 south     &  202.4630     &  47.18510     &        397&$\pm$&       123     &       3.8&$\pm$&      1.0     &      \multicolumn{3}{c}{$<$12.3} &      20.6&$\pm$&     23.6     &      \multicolumn{3}{c}{$<$0.02} &      5.64&$\pm$&     0.71     &     144.8&$\pm$&     14.5     &      18.4&$\pm$&      2.4     &     15.81&$\pm$&     0.43 \\
  M51 south     &  202.4593     &  47.18840     &        660&$\pm$&       201     &      27.0&$\pm$&      7.0     &      35.2&$\pm$&      4.2     &      64.2&$\pm$&      8.8     &      0.12&$\pm$&     0.02     &      7.60&$\pm$&     0.96     &      95.4&$\pm$&      9.5     &      15.9&$\pm$&      2.0     &     10.93&$\pm$&     0.40 \\
  M51 south     &  202.4645     &  47.18360     &        440&$\pm$&       136     &       4.0&$\pm$&      1.0     &      \multicolumn{3}{c}{$<$12.3} &      42.3&$\pm$&     18.3     &      \multicolumn{3}{c}{$<$0.03} &      3.91&$\pm$&     0.54     &     124.6&$\pm$&     12.5     &      19.3&$\pm$&      3.2     &     13.99&$\pm$&     0.40 \\
  M51 south     &  202.4686     &  47.18060     &        467&$\pm$&       144     &      10.1&$\pm$&      2.6     &      26.4&$\pm$&      4.2     &      49.8&$\pm$&     13.3     &      0.07&$\pm$&     0.02     &      4.30&$\pm$&     0.67     &     131.0&$\pm$&     13.1     &      16.7&$\pm$&      3.2     &     16.01&$\pm$&     0.60 \\
  M51 south     &  202.4573     &  47.18860     &        632&$\pm$&       192     &       4.9&$\pm$&      1.3     &      \multicolumn{3}{c}{$<$12.3} &      23.0&$\pm$&     21.7     &      \multicolumn{3}{c}{$<$0.02} &      5.03&$\pm$&     0.82     &      46.9&$\pm$&      4.7     &      19.0&$\pm$&      4.0     &     12.80&$\pm$&     1.32 \\
  \hline
\end{tabular}
\label{table:results}
\end{center}
\end{table*}

\begin{table*}[t!]
\begin{center}
\caption[h!]{Results for 3$''$ Nyquist-sampled apertures in M51.}
\setlength{\tabcolsep}{5pt} 
\begin{tabular}{lcc
r@{\hspace{0.5\tabcolsep}}C{0.2cm}@{\hspace{0.5\tabcolsep}}l
r@{\hspace{0.5\tabcolsep}}C{0.2cm}@{\hspace{0.5\tabcolsep}}l
r@{\hspace{0.5\tabcolsep}}C{0.2cm}@{\hspace{0.5\tabcolsep}}l
r@{\hspace{0.5\tabcolsep}}C{0.2cm}@{\hspace{0.5\tabcolsep}}l
R{0.8cm}@{\hspace{0.5\tabcolsep}}C{0.2cm}@{\hspace{0.5\tabcolsep}}L{0.8cm}
r@{\hspace{0.5\tabcolsep}}C{0.2cm}@{\hspace{0.5\tabcolsep}}l
r@{\hspace{0.5\tabcolsep}}C{0.2cm}@{\hspace{0.5\tabcolsep}}l
r@{\hspace{0.5\tabcolsep}}C{0.2cm}@{\hspace{0.5\tabcolsep}}l
r@{\hspace{0.5\tabcolsep}}C{0.2cm}@{\hspace{0.5\tabcolsep}}l}
\hline\hline
Region & RA$\rm_{J2000}$ & DEC$\rm_{J2000}$ 
& \multicolumn{3}{c}{$\Sigma_\star$} 
& \multicolumn{3}{c}{$L_\mathrm{TIR}$} 
& \multicolumn{3}{c}{$S_\nu^\mathrm{33\,GHz}$}
& \multicolumn{3}{c}{$f_{\rm T}^\mathrm{33\,GHz}$} 
& \multicolumn{3}{c}{$\Sigma_\mathrm{SFR}$} 
& \multicolumn{3}{c}{$I_\mathrm{HCN}$} 
& \multicolumn{3}{c}{$I_\mathrm{CO}$} 
& \multicolumn{3}{c}{$\sigma_\mathrm{HCN}$} 
& \multicolumn{3}{c}{$\sigma_\mathrm{CO}$} \\
 & (deg) & (deg) 
 & \multicolumn{3}{c}{($\mathrm{M_\odot / pc^2})$} 
 & \multicolumn{3}{c}{($10^6 L_\odot$)} 
 & \multicolumn{3}{c}{($\mu$Jy)}  
 & \multicolumn{3}{c}{(\%)}  
 & \multicolumn{3}{c}{(M$_\odot$/Myr/pc$^{2}$)} 
 & \multicolumn{3}{c}{(K\,km\,s$^{-1}$)} 
 & \multicolumn{3}{c}{(K\,km\,s$^{-1}$)} 
 & \multicolumn{3}{c}{(km\,s$^{-1}$)} 
 & \multicolumn{3}{c}{(km\,s$^{-1}$)} \\
 \hline
    M51 north     &  202.4751     &  47.20930     &        444&$\pm$&       137     &       4.5&$\pm$&      1.2     &      \multicolumn{3}{c}{$<$13.2}	&      47.5&$\pm$&     17.6	&      \multicolumn{3}{c}{$<$0.04}     &     \multicolumn{3}{c}{$<$4.17}     &	  58.7&$\pm$&	   5.9     &	  \multicolumn{3}{c}{...}     &	  8.77&$\pm$&	  0.45 \\
  M51 north     &  202.4751     &  47.21080     &        471&$\pm$&       145     &       3.5&$\pm$&      0.9     &      \multicolumn{3}{c}{$<$13.4}	&      30.3&$\pm$&     25.9	&      \multicolumn{3}{c}{$<$0.02}     &     \multicolumn{3}{c}{$<$3.87}     &	  25.7&$\pm$&	   2.6     &	  \multicolumn{3}{c}{...}     &	 10.40&$\pm$&	  1.27 \\
  M51 north     &  202.4751     &  47.21150     &        448&$\pm$&       138     &       2.8&$\pm$&      0.7     &      \multicolumn{3}{c}{$<$13.6}	&      51.3&$\pm$&     20.9	&      \multicolumn{3}{c}{$<$0.04}     &     \multicolumn{3}{c}{$<$3.87}     &	  10.4&$\pm$&	   1.0     &	  \multicolumn{3}{c}{...}     &	  7.94&$\pm$&	  1.79 \\
  M51 north     &  202.4745     &  47.20860     &        480&$\pm$&       147     &       5.6&$\pm$&      1.4     &      \multicolumn{3}{c}{$<$13.1}	&      43.1&$\pm$&     17.6	&      \multicolumn{3}{c}{$<$0.03}     &     \multicolumn{3}{c}{$<$3.39}     &	  42.9&$\pm$&	   4.3     &	  \multicolumn{3}{c}{...}     &	  8.48&$\pm$&	  0.59 \\
  M51 north     &  202.4748     &  47.20900     &        449&$\pm$&       138     &       4.7&$\pm$&      1.2     &      \multicolumn{3}{c}{$<$13.2}	&      45.4&$\pm$&     17.5	&      \multicolumn{3}{c}{$<$0.03}     &     \multicolumn{3}{c}{$<$3.81}     &	  58.4&$\pm$&	   5.8     &	  \multicolumn{3}{c}{...}     &	  9.41&$\pm$&	  0.49 \\
  M51 north     &  202.4745     &  47.20930     &        426&$\pm$&       131     &       4.1&$\pm$&      1.1     &      \multicolumn{3}{c}{$<$13.2}	&      43.9&$\pm$&     18.3	&      \multicolumn{3}{c}{$<$0.03}     &     \multicolumn{3}{c}{$<$3.90}     &	  59.3&$\pm$&	   5.9     &	  \multicolumn{3}{c}{...}     &	  9.31&$\pm$&	  0.46 \\
  M51 north     &  202.4748     &  47.20970     &        457&$\pm$&       140     &       3.9&$\pm$&      1.0     &      \multicolumn{3}{c}{$<$13.3}	&      44.7&$\pm$&     19.0	&      \multicolumn{3}{c}{$<$0.04}     &     \multicolumn{3}{c}{$<$4.41}     &	  52.4&$\pm$&	   5.2     &	  \multicolumn{3}{c}{...}     &	  8.02&$\pm$&	  0.43 \\
  M51 north     &  202.4745     &  47.21010     &        477&$\pm$&       146     &       3.4&$\pm$&      0.9     &      \multicolumn{3}{c}{$<$13.3}	&      33.8&$\pm$&     22.8	&      \multicolumn{3}{c}{$<$0.02}     &     \multicolumn{3}{c}{$<$3.87}     &	  45.1&$\pm$&	   4.5     &	  \multicolumn{3}{c}{...}     &	  8.41&$\pm$&	  0.57 \\
  M51 north     &  202.4748     &  47.21040     &        492&$\pm$&       151     &       3.5&$\pm$&      0.9     &      \multicolumn{3}{c}{$<$13.4}	&      27.6&$\pm$&     25.4	&      \multicolumn{3}{c}{$<$0.02}     &     \multicolumn{3}{c}{$<$4.02}     &	  39.9&$\pm$&	   4.0     &	  \multicolumn{3}{c}{...}     &	  8.91&$\pm$&	  0.70 \\
  M51 north     &  202.4745     &  47.21080     &        475&$\pm$&       146     &       3.1&$\pm$&      0.8     &      \multicolumn{3}{c}{$<$13.4}	&      40.0&$\pm$&     22.2	&      \multicolumn{3}{c}{$<$0.03}     &     \multicolumn{3}{c}{$<$3.30}     &	  32.1&$\pm$&	   3.2     &	  \multicolumn{3}{c}{...}     &	  9.37&$\pm$&	  0.94 \\
  ...     &  ...     &  ...     &       \multicolumn{3}{c}{...}	     &       \multicolumn{3}{c}{...}	     &      \multicolumn{3}{c}{...}	&      \multicolumn{3}{c}{...}		&      \multicolumn{3}{c}{...}     &     \multicolumn{3}{c}{...}     &	  \multicolumn{3}{c}{...}	     &	  \multicolumn{3}{c}{...}     &	  \multicolumn{3}{c}{...}	 \\
  \hline
\end{tabular}
\label{table:results2}
\end{center}
\tablefoot{The first ten rows of the table are shown for guidance on the format. This table is available in its entirety in machine-readable form.}
\end{table*}

\end{landscape}

\clearpage
\twocolumn

\end{document}